\documentclass[onecolumn,
 showpacs,amsmath,amssymb]{revtex4}
\usepackage{graphicx}
\usepackage{epstopdf}
\usepackage{bm}
\usepackage{subfigure}
\usepackage[latin1]{inputenc}
\usepackage{rotating}
\usepackage{longtable}
\usepackage{float}
\usepackage[normalem]{ulem}
\usepackage{dcolumn}% Align table columns on decimal point
\usepackage{bm}% bold math
\usepackage{color}
\usepackage{footnote}

\begin{document}

\preprint{APS/123-QED}

\title{Coherent Population Trapping Resonances in Buffer Gas-filled Cs Vapor Cells with Push-Pull Optical Pumping}

\author{Xiaochi Liu$^1$, Jean-Marc M\'erolla$^1$, St\'ephane Gu\'erandel$^2$, Christophe Gorecki$^1$, Emeric de Clercq$^2$ and Rodolphe Boudot$^1$}

\affiliation{$^1$FEMTO-ST, CNRS, 32 avenue de l'observatoire 25044 Besancon Cedex, France \\ $^2$LNE-SYRTE, Observatoire de Paris, CNRS, UPMC, 61 avenue de l'Observatoire, 75014 Paris, France}

\date{\today}

%\textcolor{blue}{}

\begin{abstract}
We report on a theoretical study and experimental characterization of coherent population trapping (CPT) resonances in buffer gas-filled vapor cells with push-pull
optical pumping (PPOP) on Cs $D_1$ line. We point out that the push-pull interaction scheme is identical to the so-called \emph{lin$\perp$lin} polarization
scheme. Expressions of the relevant dark states, as well as of absorption, are reported. The experimental setup
is based on the combination of a distributed feedback (DFB) diode laser, a pigtailed intensity Mach-Zehnder electro-optic modulator (MZ EOM) for optical sidebands
generation and a Michelson-like interferometer. A microwave technique to stabilize the transfer function operating point of the MZ EOM is implemented for proper
operation. A CPT resonance contrast as high as 78 \% is reported in a cm-scale cell for the magnetic-field insensitive clock transition. The impact of the laser
intensity on the CPT clock signal key parameters (linewidth - contrast - linewidth/contrast ratio) is reported for three different cells with various dimensions and
buffer gas contents. The potential of the PPOP technique for the development of high-performance atomic vapor cell clocks is discussed.
\end{abstract}


\maketitle

\section{Introduction}
In Coherent Population Trapping (CPT) physics \cite{Alzetta:1976, Arimondo:1996}, atoms are trapped through a destructive quantum interference process into a non-interacting coherent superposition of two long-lived ground state hyperfine levels. Under ideal conditions (isolated three-level scheme, no ground-state decoherence), this so-called dark state, obtained by simultaneous action of two resonant optical fields exactly frequency-split by the atomic ground state hyperfine frequency, is fully decoupled from the excited state. In this particular state is observed through electromagnetically induced transparency (EIT) \cite{Fleischauer:RMP:05} a narrow peak in optical transmission. The CPT resonance linewidth is limited by the atom-light interaction time and can be measured to be of a few tens to hundreds of Hz \cite{Wynands:1999, Merimaa:2003}. CPT resonance lineshapes and properties have been widely investigated for precision sensing in view of applications in high-resolution spectroscopy, magnetometers \cite{Nagel:EPL:1998, Schwindt:APL:2004}, laser cooling \cite{Aspect:JOSAB:1989}, slow light \cite{Bajcsy:Nature:2003} or quantum frequency standards \cite{Vanier:APB:2005}.\\

The short-term frequency stability $\sigma_y(\tau)$ of an atomic clock, described by the Allan deviation \cite{Rutman}, is given by:
\begin{equation}
\sigma_y(\tau) = \frac{\Delta \nu }{\nu_0 } \frac{1}{SNR} \tau^{-1/2}
\label{eq:stab}
\end{equation}
where $\nu_0$ is the clock frequency (9.192 631 770 GHz for the unperturbed Cs atom), $\Delta \nu$ is the clock resonance full-width at half maximum (FWHM), SNR is the clock resonance signal-to-noise ratio in a 1 Hz bandwidth and $\tau$ is the averaging time of the measurement. We define as the clock resonance contrast $C$ the ratio between the CPT signal height and the CPT signal background. A simplified expression of the signal to noise ratio, where photon shot noise is the only noise source considered, is given in \cite{ShahKitching:2010} by:
\begin{equation}
SNR = C \sqrt{\frac{P_{out}}{2 h \nu}}
\label{eq:SNR}
\end{equation}
where $P_{out}$ is the optical power at the output the cell and $h \nu$ is the energy of a single photon.\\

\begin{figure}
\centering
\includegraphics[width=0.7\linewidth]{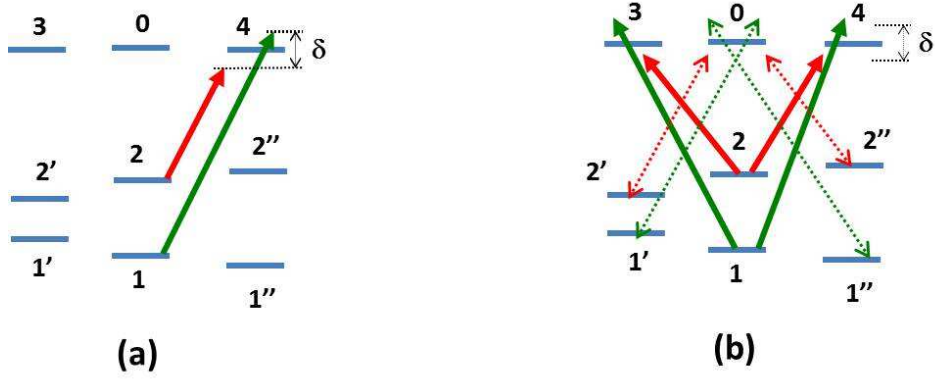}
\caption{(Color online) CPT interaction schemes. Both ground state levels are represented by 3 Zeeman sublevels (1', 1, 1'' and 2', 2, 2''). The clock transition is the 1  $\leftrightarrow$ 2 transition.
The excited state is represented by 3 Zeeman sublevels (3, 0, 4). The laser fields of amplitude $E_1$ ($E_2$) and of angular frequency $\omega_1$ ($\omega_2$) are represented by arrows issued from the (1)((2)) levels, respectively. $\delta=(\omega_1-\omega_2)-\omega_0$ is the Raman detuning, with $\omega_{0}$ the angular frequency of the clock transition. (a): $\Lambda$ scheme with circularly polarized beams ($\sigma^+$), (b): double $\Lambda$ scheme with push-pull or \emph{lin $\perp$ lin} interaction. Push-pull interaction uses two bi-chromatic laser fields circularly polarized in opposite direction ($\sigma^-$ and $\sigma^+$), one beam is phase delayed with respect to the other. In \emph{lin$\perp$lin} interaction, the two monochromatic beams are linearly polarized ($\sigma$) at right angle. Dotted lines represent two single $\Lambda$ built by $\Delta m$ =2 transitions induced by the double $\Lambda$ scheme.}
\label{fig:CPTschemes}
\end{figure}

The development of a high-frequency stability atomic clock requires to detect a narrow atomic resonance with the highest signal to noise ratio. In traditional vapor cell clocks based on CPT, atoms are probed using two circularly polarized optical fields as shown on Fig. \ref{fig:CPTschemes}(a). This interaction scheme leads numerous atoms to be concentrated at the extreme Zeeman magnetic sub-levels with the highest angular momentum $m_F$ \cite{Jau:PRL:2004-1}. Consequently, only a small fraction of atoms participate to the magnetic field insensitive clock transition resulting in CPT resonances with very low contrast of about 1 \%. Higher contrasts on the clock transition are achieved by increased laser power but at the expense of laser power broadening. Therefore, traditional CPT clocks typically operate with low laser intensities to minimize the linewidth-SNR ratio and to optimize the clock frequency stability.\\

Different optimized CPT pumping schemes were proposed to optically pump the maximum number of atoms into the dark state out of any other trap state and consequently increase dramatically the CPT resonance contrast. Y. Jau \emph{et al.} proposed the push-pull optical pumping (PPOP) scheme (see Fig. \ref{fig:CPTschemes}(b)) where atoms interact with a $D_1$ line resonant bi-chromatic optical field separated into right and left circularly polarized sub-beams, with a sub-beam delayed with  respect to the other by a half period of the hyperfine transition \cite{Jau:PRL:2004}.  The resulting light beam can be seen as alternating between the states of right and left circular polarization. The two frequencies differ by the ground state hyperfine splitting. This scheme, called a double $\Lambda$ scheme, prevents the pumping of the atoms in an  extreme Zeeman sub-level trapping state. These experiments were done on a Rb vapor cell and contrasts up to 30 $\%$ were reported. Push-pull optical pumping  was later used in an elegant mode-locked laser atomic oscillator where a K vapor cell, inserted in the external cavity of a semiconductor diode laser, acts as an optical modulator and causes the laser to be self-modulated at the clock transition frequency of the atoms \cite{Jau:PRL:2007}. Recently, Zhang et al. reported results obtained with PPOP using an original experimental set-up \cite{Zhang:OE:2012}. In 2005, T. Zanon \emph{et al.} demonstrated the detection of increased CPT resonance contrasts in Cs vapor with a so-called \emph{lin$\perp$lin} double $\Lambda$ interaction scheme using two linear orthogonally polarized optical lines \cite{Zanon:PRL:2005} (see Fig. \ref{fig:CPTschemes}(b)). CPT contrasts higher than 50 $\%$ were obtained using this technique \cite{Castagna:UFFC:2009}, allowing by combination with a Ramsey-type pulsed sequence to reach a CPT clock frequency stability better than $7 \times 10^{-13}$ at 1 s \cite{Boudot:IM:2009}. Recently, efforts have been done in different groups to realize this \emph{lin$\perp$lin} interaction scheme with a single laser source and simplified set-ups \cite{Yim:RSI:2008, Yun:RSI:2011, Yun:RSI:2012}.

Another optimized CPT pumping method based on the use of a $\sigma_+ - \sigma_-$ configuration of polarized counterpropagating waves resonant with $D_1$ line of alkali metal atoms in small-size cells was proposed in \cite{Taichenachev:JETP:2004}. This scheme is equivalent to the PPOP, but the phase condition between both beams requires a particular place of the cell along the beams and a cell length small compared to the microwave wavelength. The same group proposed the $lin||lin$ interaction scheme, using bichromatic linearly polarized light, in order to increase significantly the contrast of dark resonances on the $D_1$ line of alkali atoms with nuclear spin $I=3/2$ \cite{Taichenachev:JETP:2005}. A prototype atomic clock based on this configuration in Rb atomic vapor was developed in \cite{Mikhailov:JOSAB:2010, Zibrov:PRA:2010}. The method was also implemented with alkali atoms of large nuclear spin (Cs atom: $I=7/2$) and demonstrated a CPT constrast of about 10 $\%$ \cite{Watabe:AO:2009}. However the $lin||lin$ configuration does not create a dark state between $m_F=0$ Zeeman sublevels, but between $m_F=\pm1$ sublevels of both hyperfine levels (see section  \ref{sec:dm2}).\\

This paper reports detailed characterization of CPT resonances in buffer gas-filled Cs vapor cells using Push-Pull Optical Pumping. While most of the above-cited literature is mainly focused on the CPT clock resonance contrast, further investigations are reported here on the clock signal key parameters including linewidth, contrast and linewidth-contrast ratio for three different cells.  Section \ref{sec:theory} reports a basic theoretical analysis on the push-pull optical pumping process. The equivalence between the phase delay between two bi-chromatic circularly polarized beams and the angle between the polarization vectors of two monochromatic linearly polarized beams is shown.  It is demonstrated that PPOP and \emph{lin$\perp$lin} schemes are similar processes. Details on the dark states and absorption equations are described. Section \ref{sec:setup} describes the experimental set-up used to detect CPT resonances with PPOP on the Cs $D_1$ line at 895 nm. The key component for the generation of CPT optical sidebands is an external LiNbO$_3$ pigtailed intensity Mach-Zehnder electro-optic modulator (MZ EOM). An original microwave stabilization technique of the MZ EOM transfer function operating point is implemented and described. Section \ref{sec:expResults} reports experimental results including main characteristics of the CPT resonance lineshape for three different cells versus the laser intensity. Contrasts up to 78 $\%$ are reported in a centimetre-sized Cs vapor cell. Section \ref{sec:disc} discusses about the potential of the PPOP technique for the development of high-performance CPT atomic clocks.

\section{Theoretical analysis}
\label{sec:theory}

\subsection{Pumping waves}

The principle of push-pull CPT interaction is shown on Fig. \ref{fig:CPTschemes}(b). Push-pull interaction scheme uses two sets of circularly polarized fields, rotating in opposite directions, propagating along the z axis. Each set contains two frequencies $\omega_1$ and  $\omega_2$; one set is phase delayed with respect to the other one. The right-handed circularly polarized beam, $\overrightarrow{E_r}$, and the left-handed circularly polarized one $\overrightarrow{E_l}$, can be written using the Jones vectors \cite{a}:

\begin{eqnarray}
\left\lbrace
\begin{split}
&\overrightarrow{E_r}(t)=
\begin{pmatrix}
E_{rx}(t)\\
E_{ry}(t)\\
\end{pmatrix}
=\frac{E_1}{2} e^{i(k_1z-\omega_1t)}
\begin{pmatrix}
1\\
-i\\
\end{pmatrix}
+\frac{E_2}{2} e^{i(k_2z-\omega_2t+\eta)}
\begin{pmatrix}
1\\
-i\\
\end{pmatrix},\\
&\overrightarrow{E_l}(t)=
\frac{E_1}{2} e^{i(k_1z-\omega_1t+\varphi)}
\begin{pmatrix}
1\\
+i\\
\end{pmatrix}
+\frac{E_2}{2} e^{i(k_2z-\omega_2t+\eta+\varphi')}
\begin{pmatrix}
1\\
+i\\
\end{pmatrix},\\
\end{split}\right.
\label{Er}
\end{eqnarray}

where $E_1,E_2$ are the amplitudes of the rotating fields, $k_1$ ($k_2$) is the wave number of the field $E_1$ ($E_2$) and $\eta$ is the phase difference between the two fields. $\varphi,\varphi'$ are the phase shifts due to the optical path difference $\Delta z$ between the  $E_l$ and the $E_r$ fields such as:
\begin{equation}
\varphi=k_1\Delta z, \hspace{0.5cm} \varphi'=k_2\Delta z.
\label{phi}
\end{equation}

As the two optical frequencies $\omega_1, \omega_2$ are very close, a beat-note occurs for each beam. For the sake of simplicity the calculation is performed in the case of equal amplitudes $E_2=E_1$. Equations (\ref{Er}),(\ref{phi}) can be rewritten as :
\begin{eqnarray}
\left\lbrace
\begin{split}
&\overrightarrow{E_r}(t)=E_1 e^{\frac{i}{2}\left[(k_1+k_2)z-(\omega_1+\omega_2)t+\eta \right] )}\cos\left[\frac{(k_1-k_2)z-(\omega_1-\omega_2)t-\eta}{2}\right]\begin{pmatrix}
1\\
-i\\
\end{pmatrix},\\
&\overrightarrow{E_l}(t)=E_1 e^{\frac{i}{2}\left[(k_1+k_2)z-(\omega_1+\omega_2)t+\eta+\varphi+\varphi' \right] )}\cos\left[\frac{(k_1-k_2)z-(\omega_1-\omega_2)t-\eta+\varphi-\varphi'}{2}\right]\begin{pmatrix}
1\\
+i\\
\end{pmatrix}.
\end{split}\right.
\label{Ebatt}
\end{eqnarray}

The two sets of circularly polarized waves exhibit a beat note at the angular frequency $(\omega_1-\omega_2)/2$, phase shifted relative to each-other by:
\begin{equation}
 \theta=\dfrac{\varphi-\varphi'}{2}=\dfrac{(k_1-k_2)\Delta z}{2}=\pi\dfrac{\Delta z}{\lambda_{mw}},
 \label{phbatt}
\end{equation}
 with  $\lambda_{mw}$ the wavelength of the
microwave wave of angular frequency $(\omega_1-\omega_2)$.\\

When the terms of same frequency are grouped together, we get:

\begin{eqnarray}
\left\lbrace
\begin{split}
&\overrightarrow{E_1}=\frac{E_1}{2} e^{i(k_1z-\omega_1t)}
\left[
 \begin{pmatrix}
1\\
-i\\
\end{pmatrix}
+e^{i\varphi}
\begin{pmatrix}
1\\
+i\\
\end{pmatrix}
\right]
=E_1 e^{i(k_1z-\omega_1t+\frac{\varphi}{2})}
 \begin{pmatrix}
 \cos(\frac{\varphi}{2})\\
-\sin(\frac{\varphi}{2})\\
\end{pmatrix},\\
&\overrightarrow{E_2}=\frac{E_2}{2} e^{i(k_2z-\omega_2t+\eta)}
\left[
 \begin{pmatrix}
1\\
-i\\
\end{pmatrix}
+e^{+i\varphi'}
\begin{pmatrix}
1\\
+i\\
\end{pmatrix}
\right]
=E_2 e^{i(k_2z-\omega_2t+\eta+\frac{\varphi'}{2})}
 \begin{pmatrix}
 \cos(\frac{\varphi'}{2})\\
-\sin(\frac{\varphi'}{2})\\
\end{pmatrix}.
\end{split}\right.
\label{Elin}
\end{eqnarray}

Equations (\ref{Elin}) represent linearly polarized waves of angle $-\varphi/2$ and $-\varphi'/2$, respectively, with the $x$ axis. The angle between both polarization vectors is:
\begin{equation}
\theta=\frac{-\varphi'+\varphi}{2}=(k_1-k_2)\frac{\Delta z}{2}.
\label{theta}
\end{equation}

Then

\begin{eqnarray}
\left\lbrace
\begin{split}
&\theta=0 \hspace{0.2cm} when\ \Delta z= 2p\lambda_{mw},\\
&\theta=\frac{\pi}{2} \hspace{0.2cm} when\  \Delta z= \left( 2p+\frac{1}{2}\right)
 \lambda_{mw},
\end{split}\right.
\label{condition}
\end{eqnarray}
with $p$ an integer.\\

Here, we demonstrate that the push-pull scheme is identical to two linearly polarized waves of angular frequencies $\omega_1$ and $\omega_2$ respectively. Their polarization vectors form an angle $\theta=\pi\Delta z/\lambda_{mw}$, depending on the path difference $\Delta z$. If the path difference is realized in a Michelson interferometer like in \cite{Jau:PRL:2004}, it is worth to note that a displacement $\delta z$ of a mirror in one arm of the interferometer yields a path difference $2\delta z$. Then, the angle between polarization vectors increases from $0$ to $\pi/2$ for a displacement $\delta z = \lambda_{mw}/4$. This angle is equal to the phase shift between the two beat-notes of eq. (\ref{Ebatt}), which are in phase and in quadrature when $\theta=0$ and $\theta=\pi/2$, respectively. Finally, we point out that the push-pull scheme is identical to the \emph{lin$\perp$lin} scheme of \cite{Zanon:PRL:2005} when $\theta=\pi/2$.

\subsection{Dark States}
\subsubsection{Interaction Hamiltonian}

As the push-pull scheme is equivalent to two linear polarized fields with an angle $\theta$ between them, we consider now this last case. The same calculation can be performed with circular polarizations. Let's assume two waves copropagating along the $z$ axis, $\overrightarrow{E_1}$  parallel to $x$ axis, and $\overrightarrow{E_2}$ forming an angle $\theta$ with the $x$ axis.

\begin{eqnarray}
\left\lbrace
\begin{split}
&\overrightarrow{E_1}(t)=\mathcal{E}_1(t)\hat{e}_x=\frac{\mathcal{E}_1(t)}{\sqrt{2}}(\hat{e}_{-}-\hat{e}_+),\\
&\overrightarrow{E_2}(t)=\mathcal{E}_2(t)\left( \cos(\theta)\hat{e}_x+\sin(\theta)\hat{e}_y\right)=\frac{\mathcal{E}_2(t)}{\sqrt{2}}\left[e^{i\theta}\hat{e}_{-}- e^{-i\theta}\hat{e}_+\right],\\
\end{split}\right.
\label{defE}
\end{eqnarray}
with $\mathcal{E}_1(t)=E_1(e^{-i\omega_1 t}+e^{i\omega_1 t})/2$,
$\mathcal{E}_2(t)=E_2(e^{-i(\omega_2 t-\eta)}+e^{i(\omega_2 t-\eta)})/2$.
We used the standard basis, defined as: $\hat{e}_{\pm}=\mp(\hat{e}_x\pm i\hat{e}_y)/\sqrt{2}$, $\hat{e}_0=\hat{e}_z$.
The interaction Hamiltonian between the atoms and the electromagnetic fields is the dipole interaction Hamiltonian:
\begin{equation}
V=-\overrightarrow{D}\cdot\left( \overrightarrow{E_1}(t)+\overrightarrow{E_2}(t)\right) =V_1+V_2,
\label{V}
\end{equation}
where $\overrightarrow{D}$ is the electric dipole operator. The operators $V_1, V_2$ are equal to:

\begin{eqnarray}
\left\lbrace
\begin{split}
&V_1=\frac{\mathcal{E}_1(t)}{\sqrt{2}}\left(D_{+}-D_{-} \right) ,\\
&V_2=\frac{\mathcal{E}_2(t)}{\sqrt{2}}\left(D_{+}e^{-i\theta}-D_{-}e^{i\theta}\right).\\
\end{split}\right.
\label{Vi}
\end{eqnarray}
where $D_+$ and $D_-$ are the standard components of $D$ such as: $D_+=-(D_x+iD_y)/\sqrt{2}$, $D_-=(D_x-iD_y)/\sqrt{2}$.

We consider a four-level atom, two hyperfine levels of the ground state with quantum magnetic number $m=0$, $\vert1\rangle$ and $\vert2\rangle$, and two Zeeman sublevels  $\vert3\rangle$ and  $\vert4\rangle$ in the excited state, with $m=-1$ and $m=+1$ respectively (see Fig.\ref{fig:CPTschemes}(b)). The field $\overrightarrow{E_1}(t)$  is resonant with the transitions $\vert1\rangle- \vert3\rangle$ and $\vert1\rangle- \vert4\rangle$, $\overrightarrow{E_2}(t)$ with the transitions $\vert2\rangle- \vert3\rangle$ and $\vert2\rangle- \vert4\rangle$. The $V_2$ term can be written:

\begin{equation}
V_2=\frac{d\mathcal{E}_2(t)}{\sqrt{2}}\left[e^{-i\theta}\left( c_{42}\vert4\rangle \langle2\vert - c_{32}\vert2\rangle \langle3\vert\right)-e^{i\theta}\left(c_{32}\vert3\rangle \langle2\vert -c_{42}\vert2\rangle \langle4\vert    \right) \right] ,
\label{V2}
\end{equation}
with $d=\langle J_e\vert\vert D\vert\vert J_g\rangle/\sqrt{2J_e+1}$,  $J_e,J_g$ the total electronic angular momentum quantum numbers of the excited and ground states, respectively. The  $c_{ij}$ coefficients are the Clebsch-Gordan coefficients such that $\langle i\vert D\vert j\rangle=d_{ij}=c_{ij}d$; their values for alkali-metal atoms are given in appendix 4E of \cite{VA}.
In the rotating frame $V_2$ becomes $\widetilde{V}_2 $ \cite{Cohen:1991} :
\begin{equation}
\widetilde{V}_2=T(t)V_2T^\dagger(t) ,
\label{V2t}
\end{equation}
with:
\begin{eqnarray}
\left\lbrace
\begin{split}
&T(t)=e^{-iS(t)} ,\\
&S(t)=\omega_2 t\vert2\rangle \langle2\vert.
\end{split}\right.
\label{T}
\end{eqnarray}

In the rotating wave approximation we get:
\begin{equation}
\widetilde{V}_2=\frac{dE_2}{2\sqrt{2}}\left\lbrace e^{i(\eta-\theta)}c_{42}\vert4\rangle \langle2\vert -e^{i(\eta+\theta)}c_{32}\vert3\rangle \langle2\vert + e^{-i(\eta-\theta)}c_{42}\vert2\rangle \langle4\vert-e^{-i(\eta+\theta)}c_{32}\vert2\rangle \langle3\vert
\right\rbrace  .
\label{V2tf}
\end{equation}
The same calculation leads to:
\begin{equation}
\widetilde{V}_1=\frac{dE_1}{2\sqrt{2}}\left\lbrace c_{41}\vert4\rangle \langle1\vert -c_{31}\vert3\rangle \langle1\vert +c_{41}\vert1\rangle \langle4\vert-c_{31}\vert1\rangle \langle3\vert
\right\rbrace .
\label{V1tf}
\end{equation}

\subsubsection{Dark States}

The dark state, when it exists, is a linear superposition of the two ground states: $\vert dark\rangle=a\vert1\rangle+b\vert2\rangle$. It must satisfy:
\begin{equation}
(\widetilde{V}_1+\widetilde{V}_2)\vert dark\rangle=0.
\label{ds}
\end{equation}
Substituting $\widetilde{V}_1$ and $\widetilde{V}_2$ by their expression in (\ref{V1tf}), (\ref{V2tf}), it comes:
\begin{eqnarray}
\left\lbrace
\begin{split}
  &\frac{b}{a}=-\frac{c_{31}}{c_{32}}\frac{E_1}{E_2} e^{-i(\eta+\theta)} ,\\
  &\frac{b}{a}=-\frac{c_{41}}{c_{42}}\frac{E_1}{E_2} e^{-i(\eta -\theta)},
 \end{split} \right.
  \label{b}
\end{eqnarray}
where the first (second) equation holds for coupling with the $\vert3\rangle$ ($\vert4\rangle$) excited state, respectively. A dark state common to the two $\Lambda$ schemes exists when the following condition is satisfied:
\begin{equation}
e^{i2\theta}=\frac{c_{42}}{c_{41}}\frac{c_{31}}{c_{32}}.
\label{gammaclebsch}
\end{equation}

We consider the Cs $D_1$ line. The $\vert 1\rangle$ and $\vert 2\rangle$ states are the ground states $\vert F=3, m=0\rangle$ and $\vert F=4, m=0\rangle$, respectively, with $F$ the hyperfine quantum number. The states $\vert 3\rangle$ and $\vert 4\rangle$ are the states $\vert F', m=-1\rangle$ and $\vert F', m=+1\rangle$, respectively of the $6^2\mathrm{P}_{1/2}$ excited state hyperfine level $F'$ ($F'=3$ or $F'=4$).
The Clebsch-Gordan coefficients are such that:

\begin{eqnarray}
\left\lbrace
\begin{split}
&F'=4: \quad  c_{42}=-c_{41} \quad  c_{32}=c_{31}=c_{41}\\
&F'=3: \quad  c_{42}=-c_{41} \quad  c_{32}=c_{31}=-c_{41}.
\end{split} \right.
  \label{clebsch}
\end{eqnarray}
These relations are valid for the $D_1$ line of all alkali-metal atoms, but not for the $D_2$ line.
The dark states  corresponding to the transitions towards $\vert 3\rangle$ and towards $\vert 4\rangle$ are:
\begin{eqnarray}
\left\lbrace
\begin{split}
&\vert dark_3\rangle=\frac{1}{\sqrt{E_1^2+E_2^2}}(E_2 \vert1\rangle -E_1 e^{-i(\eta+\theta)}\vert2\rangle)\\
  &\vert dark_4\rangle=\frac{1}{\sqrt{E_1^2+E_2^2}}(E_2 \vert1\rangle +E_1 e^{-i(\eta-\theta)}\vert2\rangle).
  \end{split} \right.
  \label{dark34}
\end{eqnarray}

Note that when $\theta=0$ and $E_1=E_2$ the dark state $\vert dark_3\rangle$ for the transitions with $\vert 3\rangle$ is the bright state for the transitions with $\vert 4\rangle$, and vice-versa. In this case, no CPT effect occurs. According to (\ref{gammaclebsch}) and (\ref{clebsch}) a common dark state exists when $e^{i2\theta}=-1$, \emph{i. e.}:
\begin{equation}
\theta=\frac{\pi}{2}+k\pi,
\label{thetads}
\end{equation}
where $k$ is an integer. The common dark state is then:
\begin{eqnarray}
\left\lbrace
\begin{split}
&\vert dark_+\rangle=\frac{1}{\sqrt{E_1^2+E_2^2}}(E_2 \vert1\rangle +iE_1 e^{-i\eta}\vert2\rangle) \quad when \quad \theta = \frac{\pi}{2}+2k\pi,\\
&\vert dark_-\rangle=\frac{1}{\sqrt{E_1^2+E_2^2}}(E_2 \vert1\rangle -iE_1 e^{-i\eta}\vert2\rangle) \quad when \quad \theta = -\frac{\pi}{2}+2k\pi,\\
&i.e.,\\
&\vert dark\rangle=\frac{1}{\sqrt{E_1^2+E_2^2}}(E_2 \vert1\rangle +E_1e^{i\theta} e^{-i\eta}\vert2\rangle) \quad when \quad \theta = \frac{\pi}{2}+k\pi.\\
\end{split} \right.
\label{dark}
\end{eqnarray}
It is worth to note that, according to relations (\ref{clebsch}), this dark state is also common to transitions involving the two excited levels of the $D_1$ line of alkali-metal atoms, of hyperfine number ($F'=I-1/2$) and ($F'=I+1/2$) with $I$ the nuclear spin quantum number, as already reported in  \cite{Stahler:OL:2002}. This means that in a vapor cell where the optical transitions are broadened by Doppler effect and buffer gas collisions, the dark state built with an optical transition is not destroyed  by the coupling with the other hyperfine excited level.

\subsubsection{$\Delta m =2$ transitions}
\label{sec:dm2}

Double $\Lambda$ schemes as PPOP or \emph{lin$\perp$lin} induce two single $\Lambda$ transitions corresponding to $\Delta m =2$ \cite{Boudot:IM:2009, Knappe:2000}. These transitions are enabled by the coexistence of $\sigma^+$ and $\sigma^-$ polarized light. They are those created in $lin||lin$ configuration \cite{Taichenachev:JETP:2005}. One transition couples the states $\vert F=I-1/2,m-1\rangle$ and $\vert F=I+1/2,m+1\rangle$, and the other one the states  $\vert F=I+1/2,m-1\rangle$ and $\vert F=I-1/2,m+1\rangle$. For simplification, these states are labeled $\vert 1^\prime\rangle$, $\vert 2^{\prime\prime}\rangle$, $\vert 2^\prime\rangle$, $\vert 1^{\prime\prime}\rangle$, respectively. Calculations lead to the dark states expressions:
\begin{eqnarray}
\left\lbrace
\begin{split}
&\vert dark_{1^\prime2^{\prime\prime}}\rangle=\frac{1}{\sqrt{c_{01^\prime}^2E_1^2+c_{02^{\prime\prime}}^2E_2^2}}(c_{02^{\prime\prime}}E_2 \vert1^\prime\rangle +c_{01^\prime}E_1 e^{-i(\eta+\theta)}\vert2^{\prime\prime}\rangle)\\
&\vert dark_{2^\prime1^{\prime\prime}}\rangle=\frac{1}{\sqrt{c_{02^\prime}^2E_1^2+c_{01^{\prime\prime}}^2E_2^2}}(c_{01^{\prime\prime}}E_1 \vert2^\prime\rangle +c_{02^\prime}E_2 e^{-i(\eta-\theta)}\vert1^{\prime\prime}\rangle),\\
\end{split} \right.
\label{dm2}
\end{eqnarray}
where the $c_{ij}$ coefficients are the Clebsch-Gordan coefficients defined above. These dark states are independent and exist whatever the value of $\theta$. The corresponding CPT resonances split by increasing the magnetic field value. Unlike previous dark states, these dark states are not common to both excited hyperfine levels of the $D_1$ line. Cs atoms in these states are not fully uncoupled from the laser field because the transitions towards other excited Zeeman sublevels remain.

\subsection{Absorption}
\label{sec:absorp}

An analytical calculation of the light propagation in a medium of  four-level atoms in a double $\Lambda$ configuration has been performed by Korsunsky and Kosachiov \cite{Korsunsky:1999}. This was realized in the particular case corresponding to the $D_1$ line of an alkali-metal atom with $E_1=E_2$, at resonance. Following their calculation it is more convenient to use Rabi frequencies rather than field amplitudes, with $\Omega_{ij}=\vert d_{ij}E_{ij}\vert/\hbar$, where $E_{ij}$ is the amplitude of the field component coupling the atomic states $\vert i\rangle$ and $\vert j\rangle$. At the steady-state the absorption of the Rabi frequency along the cell length $z$ is governed by:
 \begin{equation}
\frac{d\Omega_{ij}}{dz}=-\frac{n_d}{\varepsilon_0 c\hbar}\omega_{ij}\vert d_{ij}\vert^2 \text{Im}(\widetilde{\rho}_{ij}),
\label{maxwell}
\end{equation}
where $n_d$ is the density of involved atoms, $\omega_{ij}$ is the Bohr angular frequency between $\vert i\rangle$ and $\vert j\rangle$, $\widetilde{\rho}_{ij}$ is the optical coherence in the rotating frame, $\varepsilon_0$ is the vacuum permittivity, $c$ the speed of light and $\hbar$ the Plank constant divided by $2\pi$.

From the expression of the steady-state optical coherences density matrix elements imaginary part established in \cite{Korsunsky:1999}, it can be shown that the absorption phenomenon along  $z$ can be described at low saturation $\Omega_{ij}\ll\Gamma{ij}$ by:
 \begin{equation}
\frac{d\Omega_{ij}}{dz}\simeq-\frac{n_d}{8\varepsilon_0 c\hbar}\omega_{ij}\vert d_{ij}\vert^2 \frac{\Omega_{ij}^2\left( 1-\cos(\Phi)\right) +2\Gamma_{ij}\gamma_2}{\Omega_{ij}\Gamma_{ij}},
\label{eqdif}
\end{equation}
with  $\Gamma_{ij}$ the relaxation rate of the population $\vert i\rangle$ towards $\vert j\rangle$, $\gamma_2$ the hyperfine coherence decay rate and $d_{ij}=\vert d_{ij}\vert e^{i\eta_{ij}}$. $\Phi$ is a relative phase given by:
 \begin{equation}
\Phi=(\beta_{31}-\beta_{32})-(\beta_{41}-\beta_{42}),
\label{Phi}
\end{equation}
with $\beta_{ij}=\varphi_{ij}+\eta_{ij}$ the sum of the laser phase $\varphi_{ij}$ and the phase of the atomic dipole moment $\eta_{ij}$. Using Eq. (\ref{Er}), (\ref{phi}), and (\ref{Phi}) it comes:
 \begin{equation}
(\varphi_{31}-\varphi_{32})-(\varphi_{41}-\varphi_{42})=(k_1-k_2)\Delta z.
\label{vphi}
\end{equation}
According to (\ref{clebsch}) the phases $\eta_{ij}$ are equal to $0$ or $\pi$, so that $(\eta_{31}-\eta_{32})-(\eta_{41}-\eta_{42})=\pi$. From (\ref{Phi}) and (\ref{theta}) we get:
\begin{equation}
\Phi=(k_1-k_2)\Delta z+\pi=2\pi\frac{\Delta z}{\lambda_{mw}}+\pi=2\theta+\pi,
\label{Phi0}
\end{equation}
\begin{equation}
\frac{d\Omega_{ij}}{dz}\simeq-\frac{n_d}{4\varepsilon_0 c\hbar}\omega_{ij}\vert d_{ij}\vert^2 \frac{\Omega_{ij}^2 \cos^2(\theta) +\Gamma_{ij}\gamma_2}{\Omega_{ij}\Gamma_{ij}}.
\label{eqdif1}
\end{equation}

For the $D_1$ line of alkali-metal atoms, the $\vert d_{ij}\vert$ (and the $\Gamma_{ij}$) coefficients are equal for the different involved atomic states. We note $\Gamma_{1}=\Gamma_{ij}$. With a good approximation, we consider that $\omega_{31}=\omega_{41}=\omega_{32}=\omega_{42}=\omega_{eg}$ because the optical angular frequencies $\omega_{ij}$ are very close. Consequently, we define the same absorption coefficient $\alpha_d$ for the four laser fields such as:
\begin{equation}
\alpha_d=\frac{n_d}{2\varepsilon_0 c\hbar}\frac{\omega_{eg}\vert d_{ij}\vert^2}{\Gamma_{1}}.
\label{ad}
\end{equation}

In a Cs vapor cell, most atoms are not involved in the CPT resonance because of the presence of other Zeeman sublevels for which the Raman detuning is too large. However, these atoms contribute to the laser light absorption because the related optical transitions remain at resonance. In order to take into account this additional absorption we add an absorption coefficient $\alpha_{ij}$, corresponding to the absorption of the laser beam of intensity $I_{ij}$, related to $\Omega_{ij}$ by $I_{ij}=\epsilon_0 c\hbar^2 \Omega_{ij}^2/2d^2$. $I_{ij}$ corresponds to an angular frequency $\omega_1$ or $\omega_2$ and a polarization $\sigma^-$ or $\sigma^+$. Using $I_{ij}$, $\alpha_d$, $\alpha_{ij}$ the equation (\ref{eqdif1}) becomes:
\begin{equation}
\frac{dI_{ij}}{dz}\simeq-\left(\alpha_d\cos^2(\theta)+\alpha_{ij}\right) I_{ij}-\frac{n_d}{4}\hbar\omega_{eg}\gamma_2.
\label{eqdif2}
\end{equation}
The last term of (\ref{eqdif2}) is the linear absorption expected for a closed system in a dark state ($\alpha_{ij}=0$, $\cos(\theta)=0$) \cite{Gornyi:1989,Korsunsky:1999}. The solution of (\ref{eqdif2}) is:
\begin{equation}
I_{ij}(z)\simeq \left(I_{ij}(0)+\frac{(n_d/4)\hbar\omega_{eg}\gamma_2}{\alpha_{ij}+\alpha_{d}\cos^2(\theta)} \right)  e^{-(\alpha_{ij}+\alpha_{d}\cos^2(\theta))z}-\frac{(n_d/4)\hbar\omega_{eg}\gamma_2}{\alpha_{ij}+\alpha_{d}\cos^2(\theta)},
\label{Iij}
\end{equation}
where $I_{ij}(0)$ is the beam intensity at the cell input ($z=0$), we recall that we have assumed that all the partial intensities are equal, $I_{ij}(0)=I_0/4$ with $I_0$ the full laser intensity at the cell input.\\
For reasons of symmetry the absorption coefficients $\alpha_{ij}$ of the beams $\sigma^+$ and $\sigma^+$ are equal. However, the absorption coefficients are different for the beams of angular frequencies $\omega_1$ and $\omega_2$, depending on the ground state hyperfine level involved. The full transmitted laser intensity is given by the sum of the four intensities: $I_t=\sum{I_{ij}}$.
In the usual conditions of an alkali-metal vapour cell filled with buffer gas and for usual laser intensities, we assume that $\gamma_2$ is sufficiently weak ($\Omega_{ij}^2\gg\Gamma_{ij}\gamma2$) and the transmitted total laser intensity can be simplified as follows:
\begin{equation}
I_{t}(z)\simeq \frac{I_0}{2}e^{-\alpha_{d}\cos^2(\theta)z}\left(e^{-\alpha_{1}z}+e^{-\alpha_{2}z}\right),
\label{I_t}
\end{equation}
where the subscript 1, 2 holds for the angular frequencies $\omega_1$, $\omega_2$. When $\cos(\theta)=0$, atoms involved in the dark state become fully transparent and eq. (\ref{I_t}) reduces to the well-known Beer-Lambert law.\\
Note that the model of \cite{Korsunsky:1999} is valid for a four-level atom and equal laser intensities. Here we have many levels and, even if the laser intensities are equal at the input of the cell, as the absorption coefficients due to other levels are different for the two frequencies, the laser intensities equality will not be valid after propagation inside the cell.

\section{Experimental Set-up}
\label{sec:setup}

\begin{figure}
\centering
\includegraphics[width=\linewidth]{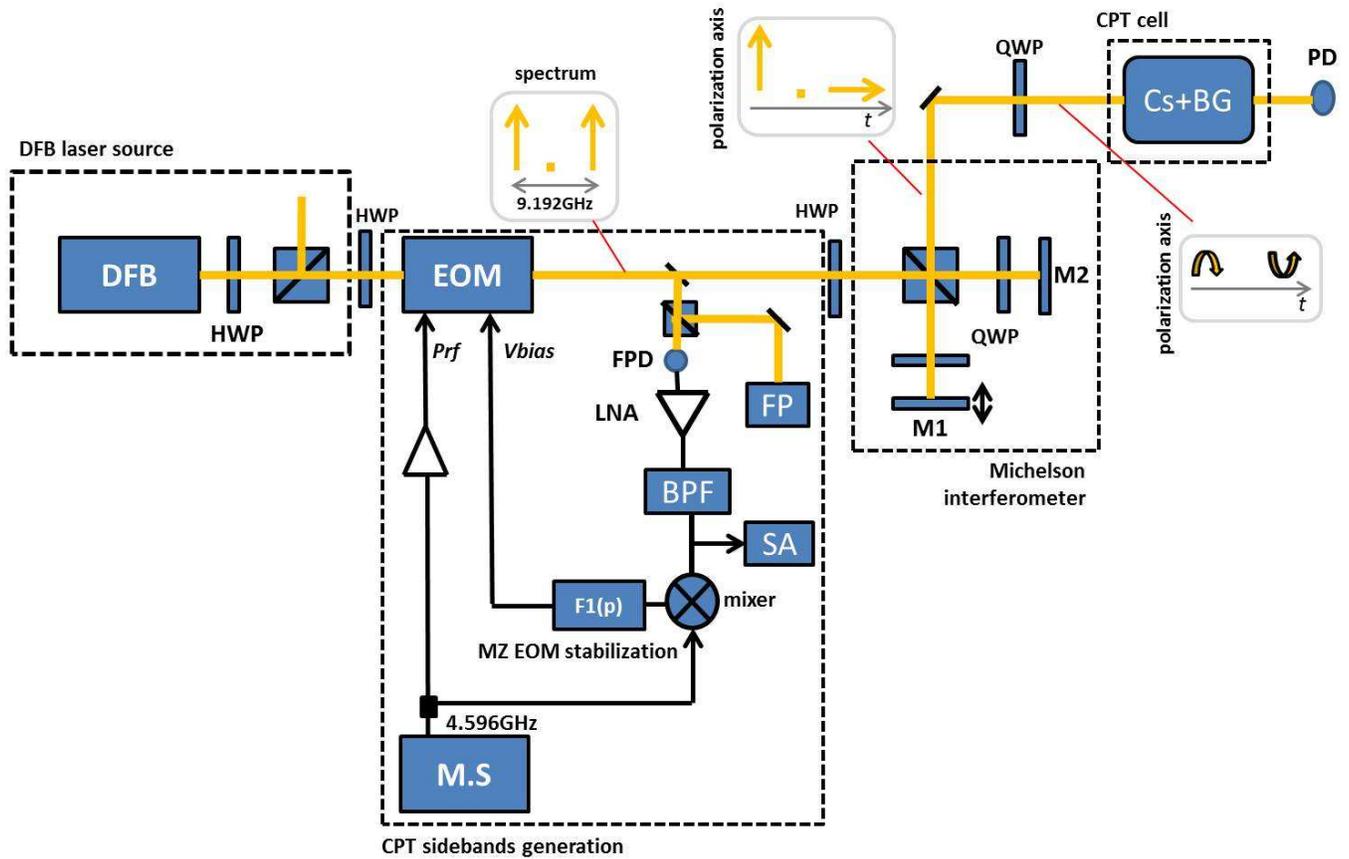}
\caption{(Color online) Experimental setup used to detect high-contrast CPT resonances in Cs vapor cells with push-pull optical pumping. A Mach-Zehnder electro-optic modulator (MZ EOM) is used to generate optical sidebands with linear and parallel polarizations, frequency-separated by the clock transition (9.192 GHz). A microwave-based stabilization technique of the transfer function operating point is implemented to stabilize the optical carrier rejection. The Michelson interferometer is used to obtain at the output of the cube C1 two time-delayed optical components with orthogonal linear polarizations. The last quarter wave plate before the cell creates alternating right and left circular polarizations for push-pull interaction. The photodiode PD detects the transmitted optical power through the vapor cell. MS: microwave synthesizer, PD: photodiode, EOM: electro-optic modulator, FPD: fast photodiode, Cs+BG: Cs + buffer gas, QWP: quarter-wave plate, HWP: half-wave plate, BPF: bandpass filter, M1 and M2: mirrors, LNA: low-noise amplifier, SA: microwave spectrum analyzer, FP: Fabry-Perot interferometer.}
\label{fig:setup}
\end{figure}

Fig. \ref{fig:setup} shows the experimental set-up used to detect high-contrast CPT resonances in Cs vapor cells using Push-Pull Optical Pumping. The setup is divided in four main basic blocks defined as the laser source, the optical sidebands generation, a Michelson interferometer to produce two orthogonally polarized beams with a phase-delay between them  \cite{Jau:PRL:2004} and the Cs vapor cell physics package. These blocks are described in detail in the following.\\

\begin{figure}[t!]
\centering
\includegraphics[width=\linewidth]{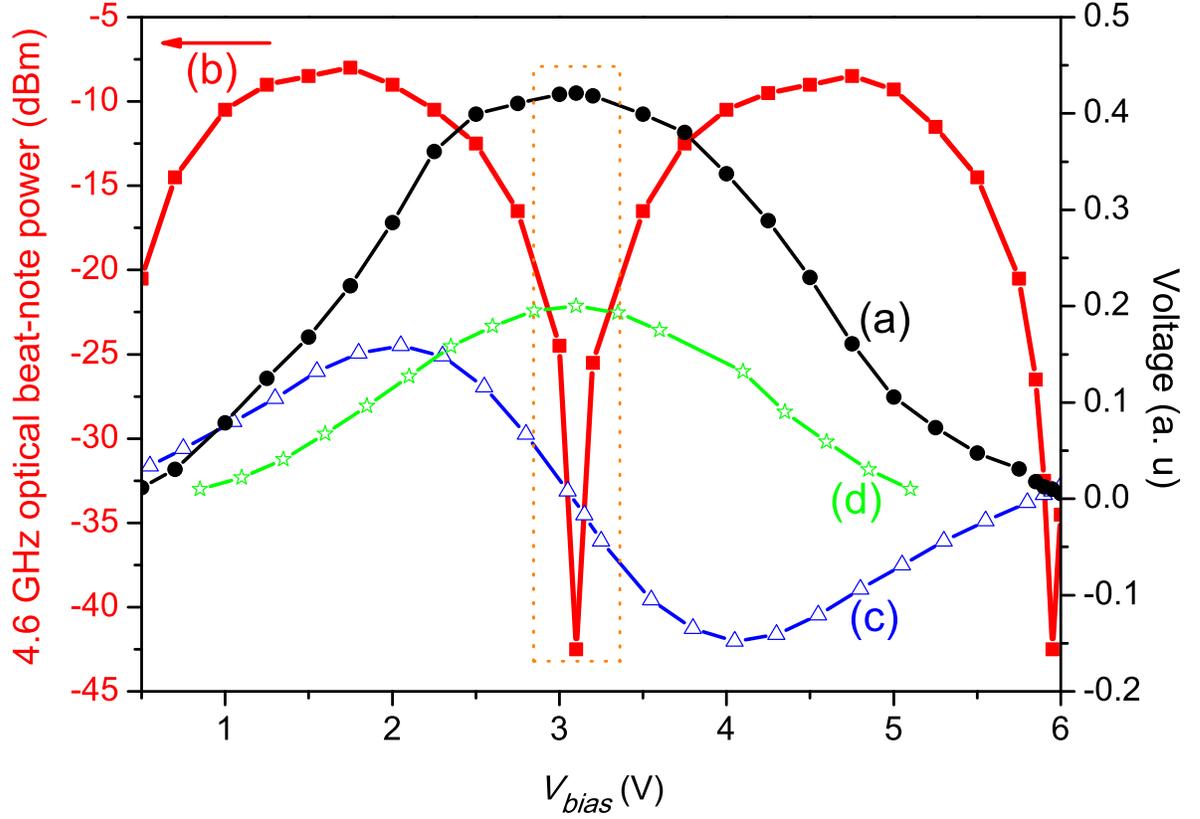}
\caption{(Color online) Impact of the EOM bias voltage $V_{bias}$ on experimental signals. The laser beam diameter is 2 cm. The cell 3 is used. (a): Optical sidebands power. (b): Power of the optically carried 4.596 GHz signal at the output of the EOM. This signal is detected by the fast photodiode and analyzed with a spectrum analyser. (c): dc signal at the output of the microwave mixer. (d): CPT signal height. } \label{fig:Signal-Vs-Bias}
\end{figure}

\begin{figure}[t!]
\centering
\includegraphics[width=\linewidth]{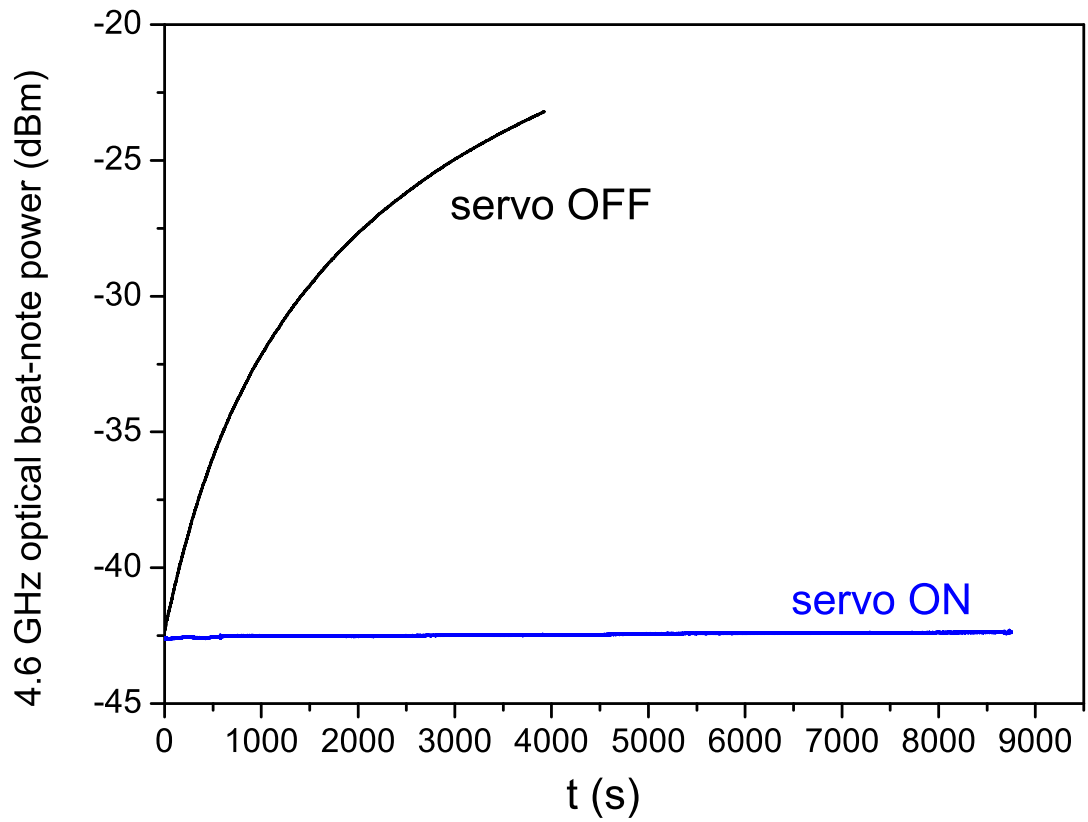}
\caption{(Color online) Power of the 4.596 GHz optically carried signal at the EOM output versus time: servo loop ON or OFF.} \label{fig:stab}
\end{figure}

The laser source is a 1MHz-linewidth DFB diode laser (Eagleyard EYP-DFB-0895) tuned on the Cs $D_1$ line at 894.6 nm \cite{Liu:IM:2012}. The laser frequency is shited fro Cs atom resonance by about 4.596 GHz. The laser light is injected into a 20 GHz bandwidth polarization maintaining (PM) pigtailed Mach-Zehnder electro-optic modulator (MZ EOM, Photline NIR-MX800-LN). CPT sidebands
are created by driving the EOM with a low noise 4.596 GHz microwave signal generated by frequency division from a 9.192 GHz frequency synthesizer \cite{Boudot:IM:NLTL:2009}. The 4.596 GHz signal power is actively stabilized by detecting a fraction of the microwave power with a Schottky diode detector and comparing the resulting output voltage to a ultra-stable voltage reference (LM399). An error signal is generated and processed in a PI controller to send a correction voltage to a voltage-controlled attenuator at the output of the synthesizer. In the locked regime, microwave power fractional fluctuations are measured to be 1 $\times$ 10$^{-3}$ at 1 s integration time and better than 1 $\times$ 10$^{-5}$ at 10000 s. Absolute phase noise performances of the synthesizer at 9.192 GHz were measured to be $-$ 65 dBc/Hz and $-$ 113 dBc/Hz at 1 Hz and 2 kHz offset frequency respectively, in good agreement with performances of frequency-multiplied state-of-the-art quartz oscillators \cite{Rubiola:xtal}. No phase noise degradation was observed between the 9.192 GHz optically carried signal at the output of the EOM and the synthesizer direct output for Fourier frequencies $f <$ 1 kHz.\\
The transfer function operating point of a MZ EOM is known to drift because of temperature variations, photorefractive effects \cite{Becker:APL:85, Harvey:JLT:98} and aging. In our experiment, the EOM is stabilized at the mK level using a temperature controller described in \cite{Boudot:RSI:2005} to a value where the EOM power transmission sensitivity to temperature variations is canceled at the first-order \cite{Liu:OL:2012}. Moreover, we implemented an original microwave technique to stabilize the operating point of the MZ EOM at the exact dark point where the optical carrier is suppressed. This method is based, as shown on Fig. \ref{fig:setup}, on the photodetection of the optically carried 4.6 GHz microwave signal, subsequent synchronous detection with a microwave mixer and active feedback control applied to the EOM dc electrode bias voltage $V_{bias}$. Fig. \ref{fig:Signal-Vs-Bias} shows versus the voltage $V_{bias}$ the first-order optical sidebands power (measured with a Fabry-Perot interferometer), the power of the 4.596 GHz optically carried signal detected at the output of the EOM with a fast photodiode, the voltage $V_m$ at the output of the mixer and the CPT signal height. For $V_{bias} \simeq$ 3.1 V, the first-order optical sidebands power (a) is maximized  while the optical carrier power is minimized. It is then clearly observed that the mixer output voltage (c) exhibits a sharp zero-crossing point that can be used for EOM stabilization. In real experience, a microwave mixer with a ultra-low intrinsic offset output voltage is finely selected to prevent to be slightly shifted from the exact dark point. For $V_{bias} \simeq$ 3.1 V, the CPT signal height (d) is maximized  while the CPT signal dc background is minimized, maximizing the CPT resonance contrast. Note that we observe that the mixer output voltage is also nulled in case where the optical sidebands power is minimum while the optical carrier is maximized. In this configuration ($V_{bias} \simeq$ 6 V in this example), the CPT signal and contrast is greatly reduced. These last results confirm the necessity to stabilize actively the EOM bias point to the exact dark point for this application. Additionally, the presence of the optical carrier could also induce in clock applications other critical issues such as light shift phenomena \cite{ZhuCutler:2000, Shah:APL:2006, Happer:APL:2009, Miletic:APB:2012}. Fig. \ref{fig:stab} reports the 4.6 GHz optical beatnote power versus time in respective cases where the servo loop is activated or not. In the free-running regime, the carrier suppression is degraded by about 17 dB in 3000 s. In the stabilized regime, the relative carrier power fluctuations are less than 5 $\times$ 10$^{-3}$ after 1000 s integration time. Relative power fluctuations at the output of the EOM are reduced by a factor 100 in the stabilized regime.\\

At the output of the EOM, the modulated light beam of fixed linear polarization passes through a half-wave plate. The incident beam is divided in two sub-beams in the two arms of a Michelson interferometer. A quarter-wave plate is placed in each arm. A fine adjustment of both quarter-wave plates and the half-wave plate allows to obtain at the output of the polarizing cube two superimposed beams of equal intensity, linearly polarized at right angle, and differing by a phase delay. A mirror, mounted on a micrometric translation stage, allows to tune the time-delay. A displacement of the mirror of $\lambda_0/4$, one quarter of the clock transition wavelength, induces a  differential optical path $\Delta\ell=\lambda_0/2$, and shifts the intensity peaks of the beat-note on one beam by half a hyperfine period with respect to the peaks of the other beam.
A last quarter-wave plate ensures that the output beams are circularly polarized, rotating in opposite directions.\\

The laser beam passes through a Cs vapor cell. The cell is temperature-stabilized at the mK level and surrounded by a static magnetic field parallel to the laser beam direction in order to lift the Zeeman degeneracy. The cell is isolated from external electromagnetic perturbations with a double-layer mu metal shield. Three different cells, named Cell 1, Cell 2 and Cell 3, were tested. Their main characteristics are reported in Table \ref{tab:cells}. The  temperature of each cell is set in order to maximize the CPT signal (see Table \ref{tab:cells}). The cell 1 is microfabricated cell realized according to the process described in \cite{Hasegawa:SA:2011} and studied in \cite{Boudot:JAP:2011} whereas cells 2 and 3 are glass-blown cm-scale pyrex vapor cells. The transmitted optical power through the cell is detected by a low noise photodiode.

\section{CPT resonances}
\label{sec:expResults}

\subsection{Push-pull CPT signal}

\begin{figure}[t!]
\centering
\includegraphics[width=\linewidth]{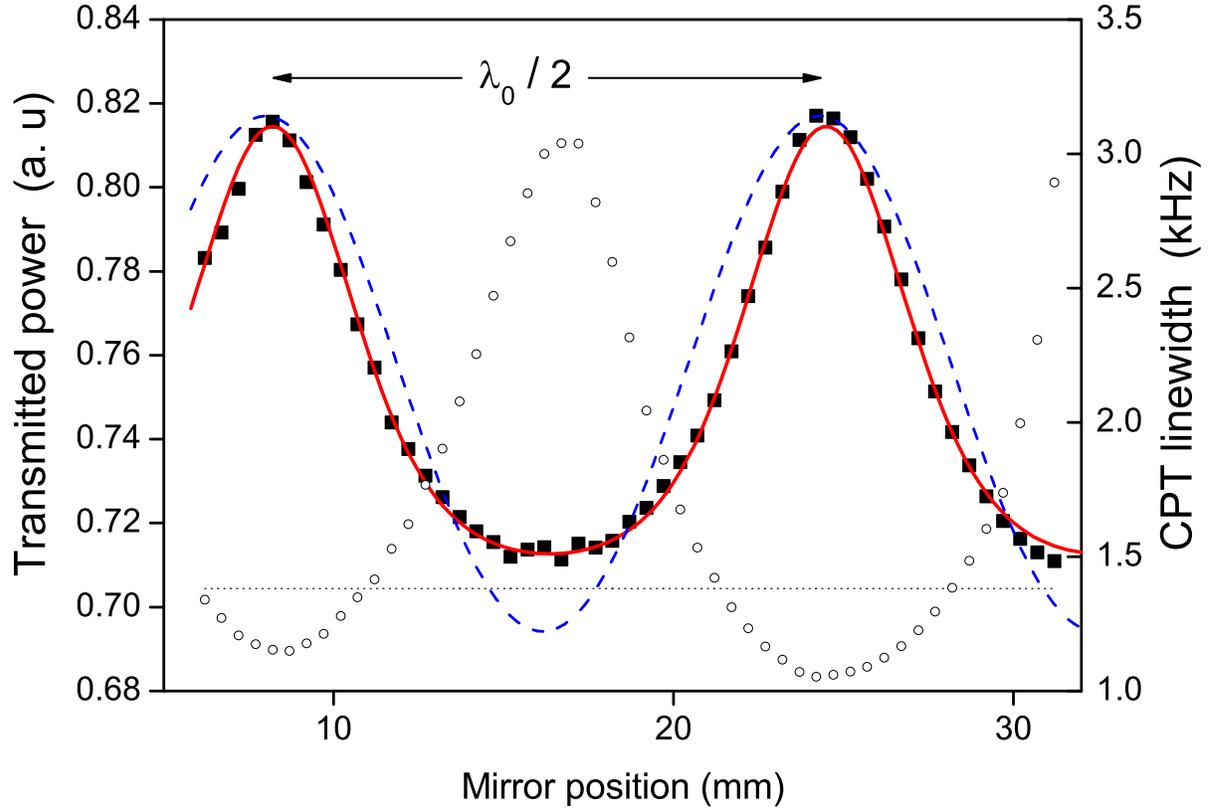}
\caption{(Color online) Transmitted  light power through cell 3 and linewidth of the CPT resonance versus the position of the mirror of one arm of the interferometer. Squares: measured light power, dotted line: light power far from CPT resonance (large Raman detuning). The dashed line is a fit of Eq. \ref{Icell}. The solid line is a fit of Eq. (\ref{Icell}) with an additional background. White dots: CPT resonance width.} \label{fig:deltaL}
\end{figure}

Fig. \ref{fig:deltaL} displays the light power transmitted through the cell 3 as a function of the position of the mirror of one arm of the Michelson interferometer. The squares are the experimental data. The dotted line is the light power far from CPT resonance for a large Raman detuning. The distance between two maxima is 16 mm which is half the wavelength of the clock transition. The optical path difference between both arms of the interferometer can be written $\Delta\ell=2(x-x_0)$, where $x$ is the mirror position and $x_0$ the position where the optical path difference is null or a multiple of $\lambda_0$ (32.6 mm).

The theoretical signal, given by Eq. (\ref{I_t}) with $\theta= 2\pi(x-x_0)/\lambda_0$, can be written:
\begin{equation}
S = S_{max} e^{-\alpha_d\cos^2(2\pi\frac{x-x0}{\lambda_{0}})L},
\label{Icell}
\end{equation}
with $S$ is the transmitted power, $S_{max}$ the signal at CPT resonance maximum and $L$ the cell length. The dashed curve is a fit of experimental data with $\alpha_d$ and $x_0$ the adjustable parameters. The agreement is not good.  The solid line on Fig.  \ref{fig:deltaL} is a fit of Eq. (\ref{Icell}) with  an additional background. Surprisingly, the agreement is found to be very good. Nevertheless, the fitted value of the background is significant and very close of the off-CPT resonance signal. We have no explanation for this and Eq.(\ref{Icell}) with a background added can be considered as a phenomenological model.\\
In our experimental conditions, the last term of Eq (\ref{Iij}) is negligible and can not explain the disagreement between the theoretical expression and experimental data. At the opposite, the latter could be explained by an oversimplification of our model developed in section \ref{sec:absorp}. Indeed, our model is based on analytical expressions from \cite{Korsunsky:1999} calculated for a closed four level scheme whereas in our experience, atoms can circulate on the 16 Zeeman sublevels of the ground state. At CPT resonance ($\theta =\pi/2$), a significant part of the atoms are pumped in the dark state. When $\theta$ deviates from $\pi/2$, a growing part of these atoms leaks towards other sublevels by absorption and spontaneous emission. Then, the atomic population of the states involved in the dark state decreases. This implies that the absorption coefficients $\alpha_d$, $\alpha_1$, $\alpha_2$  of Eq. (\ref{Iij}) are not constant but $\theta$ dependent. This could explain that experimental peaks are sharper than in the theoretical case. For further study, a complete analytical calculation does not seem possible. A full numerical calculation taking into account all the sublevels and all the optical transitions is required. The leak of the ($m=0$) clock level populations when $\theta$ deviates from $\pi/2$ corresponds to an increase of the hyperfine coherence relaxation $\gamma_2$, and consequently of the width of the CPT resonance. This is confirmed by the measurement of the resonance width as a function of $\theta$, shown by white dots curve on Fig. \ref{fig:deltaL}. It is clearly seen that the CPT width and the signal are in opposite phase.

\begin{figure}[t!]
\centering
\includegraphics[width=\linewidth]{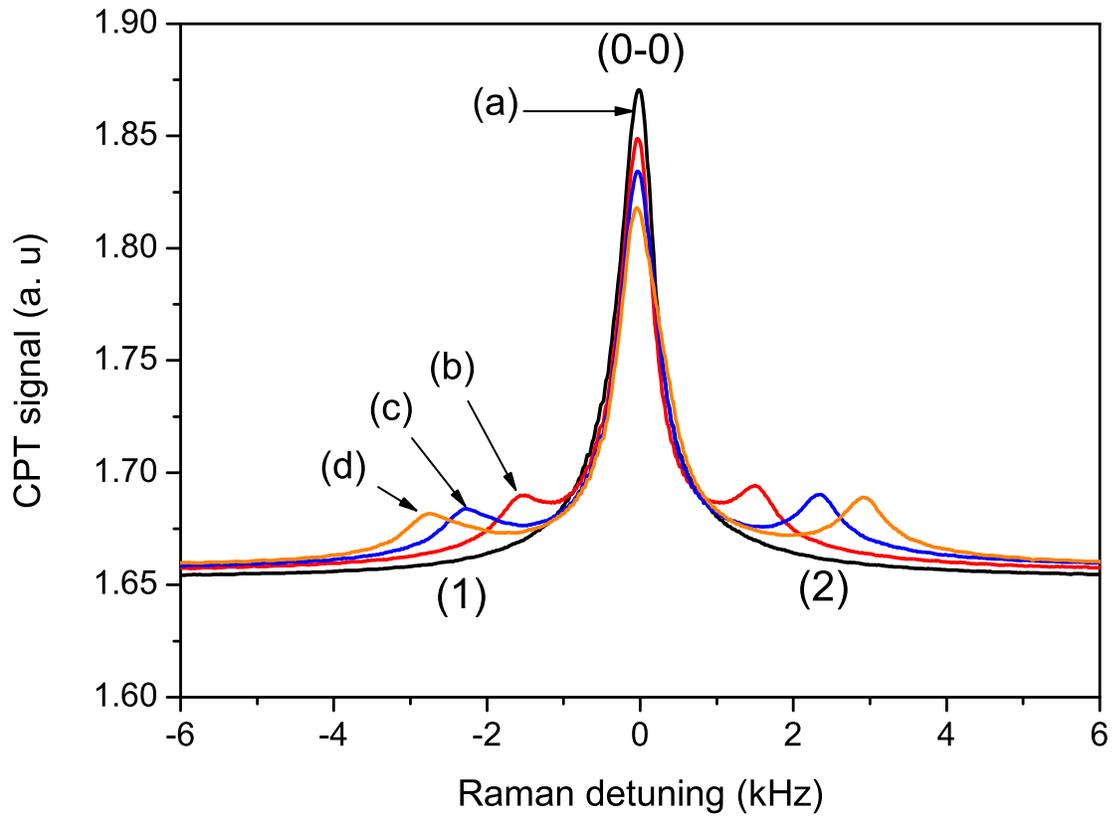}
\caption{(Color online) CPT spectra for various static magnetic fields: (a): 45 mG, (b): 1.335 G, (c): 2.07 G, (d): 2.531 G. Neighboring single $\Lambda$ $\Delta m$ = 2 transitions close to the central clock transition are noted (1) and (2). For clarity of the figure, the Zeeman frequency shift of the clock transition is compensated for each value of $B$.} \label{fig:dm2}
\end{figure}

On Fig. \ref{fig:deltaL}, the difference between minimum experimental values and the off-CPT resonance signal (dashed line) can be explained by the presence of neighboring $\Delta m$= 2 CPT resonances close to the clock transition (see section \ref{sec:dm2}). For illustration, Fig. \ref{fig:dm2} shows typical CPT spectra observed in the cell 3 for various static magnetic fields. $\Delta m$= 2 transitions connecting on Fig. \ref{fig:CPTschemes} levels $2' \rightarrow 0 \rightarrow 1''$ and $2'' \rightarrow 0 \rightarrow 1'$ are noted on Fig. \ref{fig:dm2} (1) and (2) respectively. For low static magnetic fields, adjacent peaks from $\Delta m$= 2 transitions are not resolved. When the static magnetic field is increased, resonances (1) and (2) move apart the clock transition with opposite directions. We measured a sensitivity of $-$ 1141 Hz/G + 26 Hz/G$^2$ and $+$ 1125 Hz/G + 24 Hz/G$^2$ for the frequency splitting $\nu_{(1)} - \nu_0$ and $\nu_{(2)} - \nu_0$ respectively, that is in reasonable agreement with theoretical values $\nu=\pm1116.5$ Hz/G $+ 27$ Hz/G$^2$. As expected by calculations, we observed that the amplitude of the neighboring $\Delta m$= 2 transitions resonances does not change with the mirror position. Compared to the clock transition, the relative amplitude of the $\Delta m$= 2 transitions was measured to be 6\% in the experimental conditions of Fig. \ref{fig:deltaL}. Note that the presence of these $\Delta m$= 2 transitions induces a broadening of the CPT clock resonance and can cause a supplemental sensitivity to the static magnetic field.\\

\begin{figure*}
\centering
\subfigure[]{\includegraphics[width=0.7\linewidth]{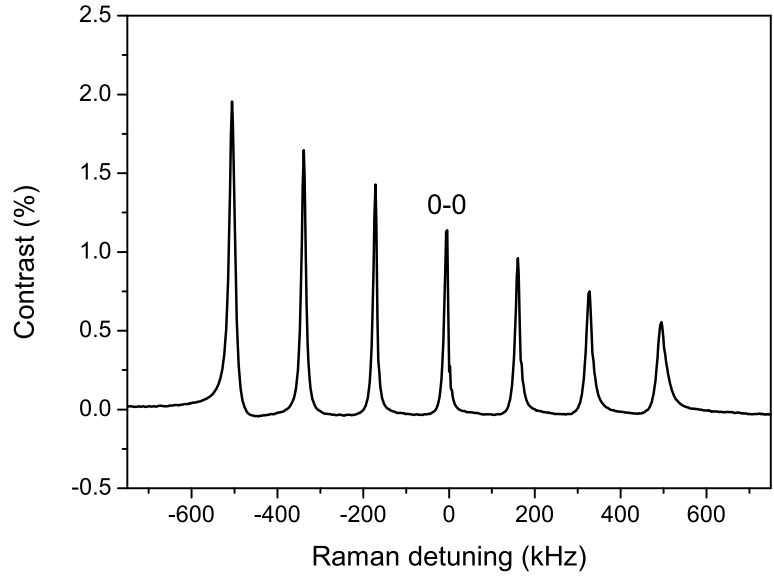}
\label{fig:Fig5a}} \vfill
\subfigure[]{\includegraphics[width=0.7\linewidth]{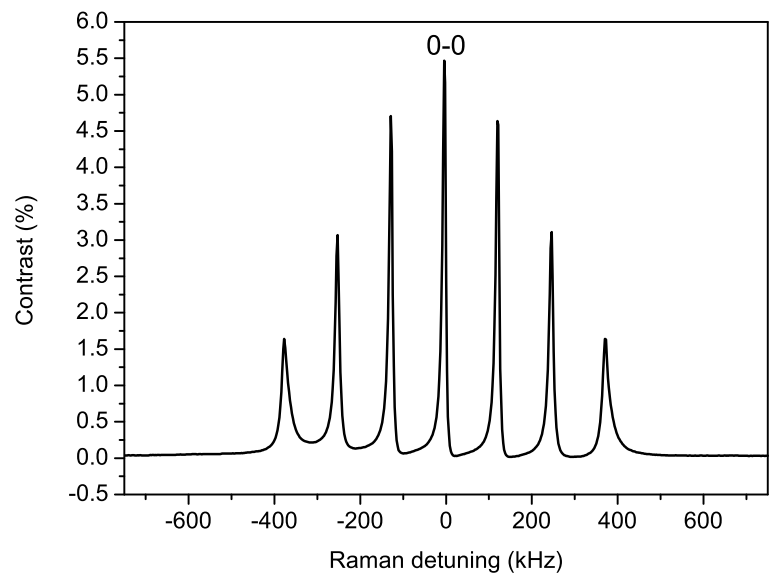}
\label{fig:Fig5b}}
 \caption{(Color online) Zeeman spectrum detected in the cell 2. The total laser power is 1.05 mW. The beam diameter is 1 cm. The static magnetic field is not the same for both curves (without consequence on the CPT signal). (a): Circular polarization (b): Push-pull optical pumping.}
\end{figure*}

\begin{figure}[t!]
\centering
\includegraphics[width=\linewidth]{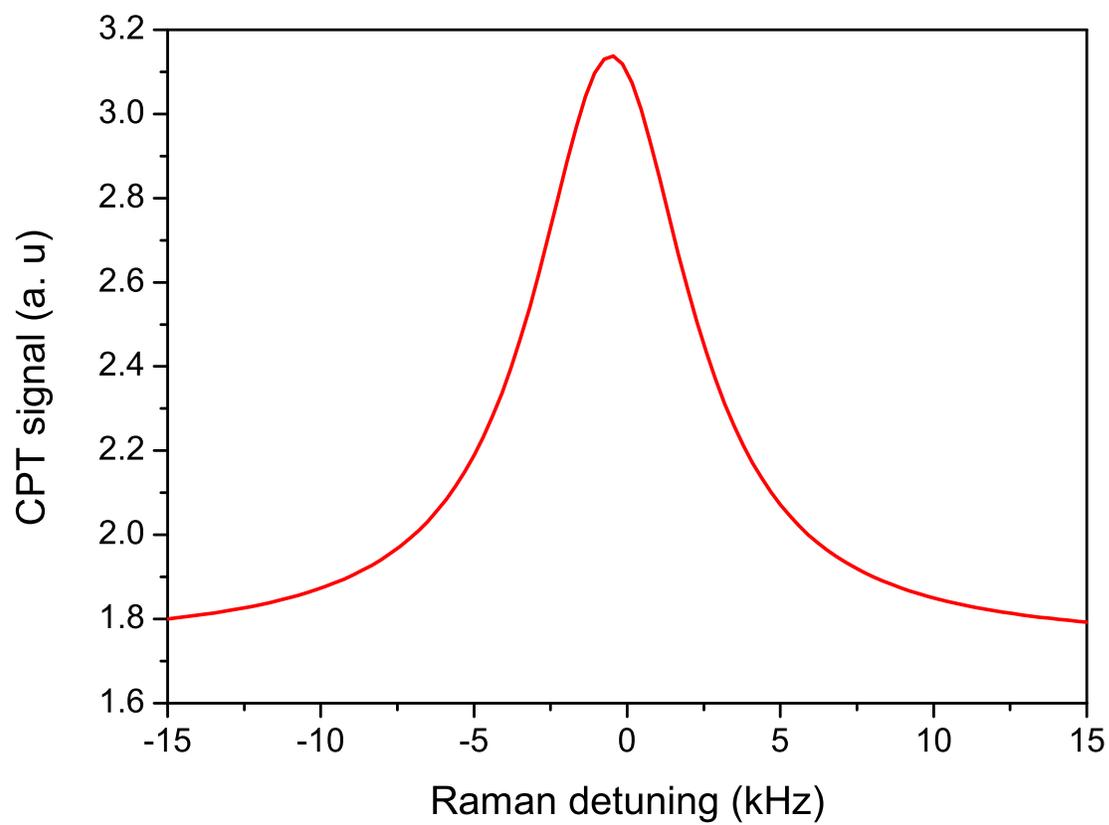}
\caption{(Color online) CPT resonance detected in the cell 3. The laser power is 3 mW. The beam diameter is 6 mm. The contrast is 78 \%.} \label{fig:C80}
\end{figure}

Figs \ref{fig:Fig5a} and \ref{fig:Fig5b} show a Zeeman spectrum obtained in the cell 2 with classical circular polarization and PPOP, respectively, for identical total incident laser intensity. It is clear that most of the atoms are lost in extreme Zeeman sublevels with circular polarization. At the opposite, the number of atoms that participate to the central clock transition is greatly favored with push-pull optical pumping. In this case, the Zeeman spectrum is found to be well symmetrical and the clock resonance contrast is here 5.5 \%.
Fig. \ref{fig:C80} shows that extremely high contrasts can be obtained in the cell 3 by operating with high laser intensity. In this case, a CPT resonance with a contrast of 78 \% is detected with a laser beam diameter of 6 mm. To our knowledge, this value is among the best contrasts ever measured on a CPT resonance.

\subsection{PPOP potential for atomic clocks}
This section is devoted to report the characterization of CPT resonances with PPOP in view of application to frequency standards. From \cite{Jau:PRL:2004}, it is noted that push-pull optical pumping operates correctly only if alternating polarizations are used and if the time-averaged optical pumping rate $\Gamma_p$ greatly exceeds the ground-state relaxation rate $\gamma_2$. This condition is fulfilled for the three cells of different size shown in Table \ref{tab:cells}.\\
Figures 9, 10, 11 display the linewidth, contrast and linewitdh-contrast ratio versus the laser intensity for the cells 1, 2 and 3 respectively. As expected and already observed for single $\Lambda$ schemes \cite{Knappe:JOSAB:2001}, the linewidth increases with higher laser intensity. For all the cells, the CPT linewidth-laser intensity dependence is well fitted by a linear function and is measured to be (0.21 $I$+ 1.32) kHz, (0.19 $I$+ 0.59) kHz and (0.058 $I$+ 0.42) kHz for the cells 1, 2 and 3 respectively ($I$ is the laser intensity in $\mu$W/mm$^2$). The resonance contrast increases with higher laser intensity because of increased optical pumping rate. This behavior is similar to the one reported in \cite{Jau:PRL:2004}. For a laser intensity of 30 $\mu$W/mm$^2$, the contrast is measured to be 4.7 \%, 10.2 \% and 45 \% for the cells 1, 2 and 3 respectively. For the cell 1 (microcell), the maximum contrast is measured to be about 6 \%. This value is only 2-3 times higher than the contrast obtained with classical circular polarization. The linewidth-to-contrast ratio is minimized for laser intensities of about 3, 5 and 30 $\mu$W/mm$^2$ for cells 1, 2 and 3 respectively.\\
The experimental values of the linewidth-to-contrast ratios allow us, by combining Equations (\ref{eq:stab}) and (\ref{eq:SNR}), to foresee the corresponding shot-noise limited short term frequency stability of an atomic clock. Results are shown on Fig. \ref{fig:stab-bilan} as a function of the laser intensity  for the different cells.
Clock resonance key parameters ($\Delta \nu$, $C$, $P_{out}$) are extracted from experimental data. We note for all cells an improvement of the clock frequency stability with increased laser intensity until a plateau after which the clock stability slightly degrades. The optimum laser intensity $I_{opt}$ for best stability performances is measured to be 1-2, 10 and 33 $\mu$W/mm$^2$ for the cells 1, 2 and 3 respectively. For the longest cell (cell 3), only taking into account photon shot noise, the expected clock short-term frequency stability is close to 2 $\times$ 10$^{-14}$ for a laser intensity of 33 $\mu$W/mm$^2$. This value is about one order of magnitude better than frequency stability performances of state-of-the-art vapor cell frequency standards \cite{Micalizio:Metrologia:2012}.

\section{Discussion}
\label{sec:disc}
CPT physics is typically of relevant interest for the development of low power consumption chip-scale atomic clocks \cite{Knappe:Elsevier:2010}. Such miniature clocks are based on microfabricated alkali vapor cells (similar to the cell 1 for example) that need to be filled with a high buffer gas pressure to obtain narrow resonances through the Dicke effect \cite{Dicke:PR:1953}. In this study, a poor efficiency gain of the PPOP technique to detect high-contrast CPT resonances was obtained in the microfabricated cell (cell 1). PPOP was found to be more efficient to obtain high contrasts when the cell dimensions are increased. This can be mainly related to two parameters. At CPT resonance, according to Eq. (\ref{Iij}), the signal decreases with increased hyperfine relaxation rate $\gamma_2$, which is one order of magnitude larger in mm-sized cells. Out of CPT resonance, the background signal is governed by the absorption coefficient which varies as the inverse of $\Gamma$, the width of the optical transition, which is also larger in small cells due to a higher buffer gas pressure. We can then expect for small cells, a smaller signal at resonance and a higher background signal out of resonance, which leads to a comparatively smaller contrast. Nevertheless a full numerical calculation would be needed to explain this behavior. In other words, this study suggests that optimized CPT pumping schemes such as PPOP or \emph{lin $\perp$ lin} can not be used efficiently to improve the performances of chip-scale atomic clocks, but are very promising for more conventional atomic clocks using cm-sized cells. Similarly to the use of circularly polarized light, we measured with PPOP that high contrasts are achieved by increased laser power at the expense of laser power broadening of the clock transition. To circumvent this issue, PPOP could be associated with a pulsed Ramsey-type interaction \cite{Castagna:UFFC:2009, Boudot:IM:2009, Zanon:PRA:2011, Yun:EPL:2011}.\\

To summarize, we reported a basic theoretical analysis on push-pull optical pumping in Cs vapor cells. Analogies between the PPOP and the \emph{lin $\perp$ lin} technique were pointed out. Conditions of dark states existence and absorption calculations were reported and detailed. We developed an experimental system combining a single diode laser, an external Mach-Zehnder electro-optic modulator and a Michelson interferometer to detect high-contrast CPT resonances in Cs vapor cells filled with buffer gas. We compared Zeeman spectra detected in a buffer gas-filled Cs vapor cell using standard circular polarization and PPOP to verify that the latter allows to pump an increased number of atoms in the 0-0 clock transition. We evaluated the impact on the CPT 0-0 resonance of the laser intensity for three different cells, including a microfabricated cell. Contrasts up to 78 \% were achieved in a 5 cm-long and 2 cm-diameter Cs-N$_2$-Ar cell. We pointed out that the efficiency of the PPOP technique to detect high-contrast CPT resonances is greatly reduced when the cell dimensions are strongly reduced. We proposed the combination of PPOP with pulsed CPT interaction for the development of high-frequency stability atomic clocks.

\section*{Acknowledgments}\label{sec:Acknow}
This work was mainly done in the frame of a scientific program contract
with Laboratoire National de M\'etrologie et d'Essais (LNE-DRST 10-3-005). A part of the study was done thanks to
Agence Nationale de la Recherche (ANR ISIMAC project, ANR-11-ASTR-0004) and D\'el\'egation G\'en\'erale de l'Armement (DGA). X. Liu is
financially supported by R\'egion de Franche-Comt\'e. The authors thank C. Rocher and P. Abb\'e (FEMTO-ST) for
technical support and electronics design and M. Hasegawa for the realization of the microfabricated cell.\\

\clearpage
\section*{List of Table Captions}
Tab. 1. Characteristics of tested buffer gas filled cells and relevant parameters. For buffer gas mixtures, $r$ is defined as the ratio $P_{Ar}/P_{N_2}$ between Ar and N$_2$ partial pressures. The experimental data are the features of the cell, the buffer gases, the laser beam, the contrast and the CPT linewidth. All other data are calculated values.  \\

\clearpage

\section*{List of Figure Captions}

Fig. 1. (Color online) CPT interaction schemes. Both ground state levels are represented by 3 Zeeman sublevels (1', 1, 1'' and 2', 2, 2''). The clock transition is the 1  $\leftrightarrow$ 2 transition.
The excited state is represented by 3 Zeeman sublevels (3, 0, 4). The laser fields of amplitude $E_1$ ($E_2$) and of angular frequency $\omega_1$ ($\omega_2$) are represented by arrows issued from the (1)((2)) levels, respectively. $\delta=(\omega_1-\omega_2)-\omega_0$ is the Raman detuning, with $\omega_{0}$ the angular frequency of the clock transition. (a): $\Lambda$ scheme with circularly polarized beams ($\sigma^+$), (b): double $\Lambda$ scheme with push-pull or \emph{lin $\perp$ lin} interaction. Push-pull interaction uses two bi-chromatic laser fields circularly polarized in opposite direction ($\sigma^-$ and $\sigma^+$), one beam is phase delayed with respect to the other. In \emph{lin$\perp$lin} interaction, the two monochromatic beams are linearly polarized ($\sigma$) at right angle. Dotted lines represent two single $\Lambda$ built by $\Delta m$ =2 transitions induced by the double $\Lambda$ scheme.\\

Fig. 2. (Color online) Experimental setup used to detect high-contrast CPT resonances in Cs vapor cells with push-pull optical pumping. A Mach-Zehnder electro-optic modulator (MZ EOM) is used to generate optical sidebands with linear and parallel polarizations, frequency-separated by the clock transition (9.192 GHz). A microwave-based stabilization technique of the transfer function operating point is implemented to stabilize the optical carrier rejection. The Michelson interferometer is used to obtain at the output of the cube C1 two time-delayed optical components with orthogonal linear polarizations. The last quarter wave plate before the cell creates right and left circular polarizations for push-pull interaction. The photodiode PD detects the transmitted optical power through the vapor cell. MS: microwave synthesizer, PD: photodiode, EOM: electro-optic modulator, FPD: fast photodiode, Cs+BG: Cs + buffer gas, QWP: quarter-wave plate, HWP: half-wave plate, BPF: bandpass filter, M1 and M2: mirrors, LNA: low-noise amplifier, SA: microwave spectrum analyzer, FP: Fabry-Perot interferometer.\\

Fig. 3. (Color online) Impact of the EOM bias voltage $V_{bias}$ on experimental signals. The laser beam diameter is 2 cm. The cell 3 is used. (a): Optical sidebands power. (b): Power of the optically carried 4.596 GHz signal at the output of the EOM. This signal is detected by the fast photodiode and analyzed with a spectrum analyser. (c): dc signal at the output of the microwave mixer. (d): CPT signal height.\\

Fig. 4. (Color online) Power of the 4.596 GHz optically carried signal at the EOM output versus time: servo loop ON or OFF.\\

Fig. 5. (Color online) Transmitted  light power through cell 3 and linewidth of the CPT resonance versus the position of the mirror of one arm of the interferometer. Squares: measured light power, dotted line: light power far from CPT resonance (large Raman detuning). The dashed line is a fit of Eq. \ref{Icell}. The solid line is a fit of Eq. (\ref{Icell}) with an additional background. White dots: CPT resonance width.\\

Fig. 6. (Color online) CPT spectra for various static magnetic fields: (a): 45 mG, (b): 1.335 G, (c): 2.07 G, (d): 2.531 G. Neighboring single $\Lambda$ $\Delta m$ = 2 transitions close to the central clock transition are noted (1) and (2). For clarity of the figure, the Zeeman frequency shift of the clock transition is compensated for each value of $B$.\\

Fig. 7. (Color online) Zeeman spectrum detected in the cell 2. The total laser power is 1.05 mW. The beam diameter is 1 cm. The static magnetic field is not the same for both curves (without consequence on the CPT signal). (a): Circular polarization (b): Push-pull optical pumping.\\

Fig. 8. (Color online) CPT resonance detected in the cell 3. The laser power is 3 mW. The beam diameter is 6 mm. The contrast is 78 \%.\\

Fig. 9. (Color online) CPT resonance linewidth (a), CPT resonance contrast (b) and linewidth/contrast ratio (c) versus laser intensity for the cell 1. The laser beam diameter is 2 mm.\\

Fig. 10. (Color online) CPT resonance linewidth (a), CPT resonance contrast (b) and linewidth/contrast ratio (c) versus laser intensity for the cell 2. The laser beam diameter is 6 mm.\\

Fig. 11. (Color online) CPT resonance linewidth (a), CPT resonance contrast (b) and linewidth/contrast ratio (c) versus laser intensity for the cell 3. The laser beam diameter is 6 mm.\\

Fig. 12. (Color online) Estimated shot-noise limited clock short-term frequency stability at 1 s for the different cells with PPOP interaction. The combination of Eq. 1 and Eq. 2 is used for the calculation. The laser beam diameter is 2 mm for the cell 1 and 6 mm for cells 2 and 3. (a): Cell 1, (b): Cell 2, (c): Cell 3.\\

%\listoffigures

\clearpage

\begin{table}
\caption{Characteristics of tested buffer gas filled cells and relevant parameters. For buffer gas mixtures, $r$ is defined as the ratio $P_{Ar}/P_{N_2}$ between Ar and N$_2$ partial pressures. The experimental data are the features of the cell, the buffer gases, the laser beam, the contrast and the CPT linewidth. All other data are calculated values. }
\label{tab:cells}\vspace*{-1ex}
\begin{center}\begin{tabular}{llll}\hline
Cell  & 1 & 2& 3 \\ \hline
Buffer gas & Ne& N$_2$-Ar & N$_2$-Ar \\
Buffer gas total pressure $P$ (Torr) & 75 & 15 & 15\\
Buffer gas ratio $r$ & - & 0.4 & 0.4\\
Cell length $L$ (cm) & 0.14 & 1 & 5\\
Cell diameter $D$ (cm) & 0.2 & 1 & 2\\
Cell temperature $T_{cell}$ ($^{\circ}$C) & 80 & 38 & 32\\
Cs atomic density  $n_{Cs}$ (at/cm$^3$) & 4.0 $\times$ 10$^{12}$ & 1.6 $\times$ 10$^{11}$ & 9.0 $\times$ 10$^{10}$ \\
Cs-wall collisions relaxation rate $\gamma_w$ (rad/s) & 2470  & 442 & 374\\
Cs-Buffer gas relaxation rate $\gamma_{bg}$ (rad/s) & 16 & 19.4 & 19.2\\
Spin-exchange relaxation rate $\gamma_{se}$ (rad/s) & 2040 & 73 & 42\\
Ground-state relaxation rate $\gamma_{2}$ (rad/s) & 4520  & 534 & 435\\
% CPT coherence lifetime $T_2 = 1/ \gamma_{2}$ (ms) & 0.22 & 1.9 & 2.3\\
CPT linewidth FWHM (kHz) & 1.4 & 0.17 & 0.14\\
Laser power $P_{L}$ (mW) & 0.5 & 0.5 & 0.5\\
Laser diameter (cm)  & 0.2 & 0.6 & 0.6\\
Rabi Frequency $\Omega$ (rad/s) & 2.9 $\times$ 10$^7$ & 9.5 $\times$ 10$^6$ & 9.5 $\times$ 10$^6$\\
Excited level relaxation rate $\Gamma^{\star}$ (rad/s) & 5.9 $\times$ 10$^9$  & 1.8 $\times$ 10$^9$ & 1.8 $\times$ 10$^9$\\
Optical transition width $\Gamma^{\star} / 2 \pi$ (MHz) & 937 & 292 & 292\\
Optical transition shift  (MHz) & -155 & -136 & -131\\
Number of Cs atoms in the laser beam  & 1.8 $\times$ 10$^{10}$ & 4.5 $\times$ 10$^{10}$ & 13
 $\times$ 10$^{10}$\\
Experimental contrast (\%)  & 6 & 10.5  & 50\\
Experimental CPT linewidth (kHz)  & 1.32 & 0.59  & 0.42\\
\hline
\end{tabular}\end{center}
\end{table}

\clearpage
%%%%%%%%%%%%%%%%%%%%%%%%%%%%%%%%%%%%%%%%%%%%%%%%%%%%% ATTENTION !!!!

%%%%%%%%%%%%%%%%%%%%%%%%%%%%%%%%%%%%%%%%%%%%%%%%%%%%% ATTENTION !!!!

\clearpage

\clearpage

\clearpage

\clearpage

\clearpage

\clearpage

\clearpage

\clearpage
\begin{figure*}
\centering
\subfigure[]{\includegraphics[width=0.4\linewidth]{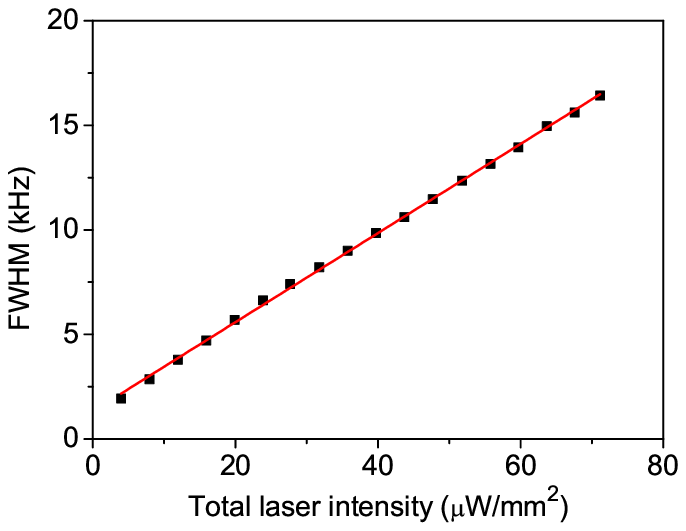}
\label{fig:largeur-microcell}} \vfill
\subfigure[]{\includegraphics[width=0.4\linewidth]{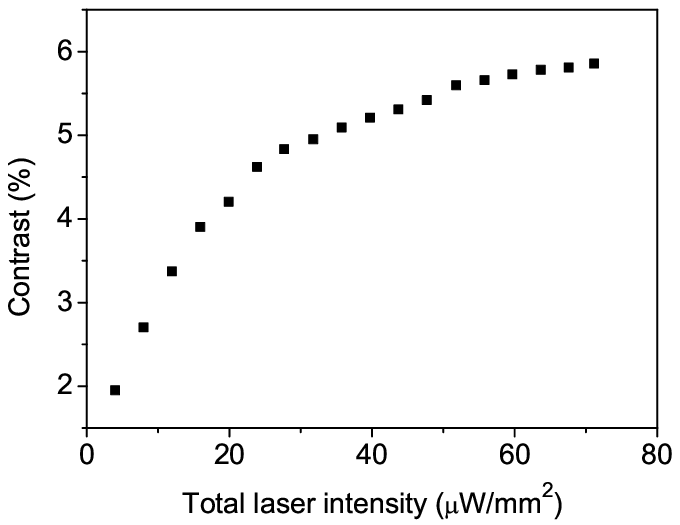}
\label{fig:C-microcell}}\vfill
\subfigure[]{\includegraphics[width=0.4\linewidth]{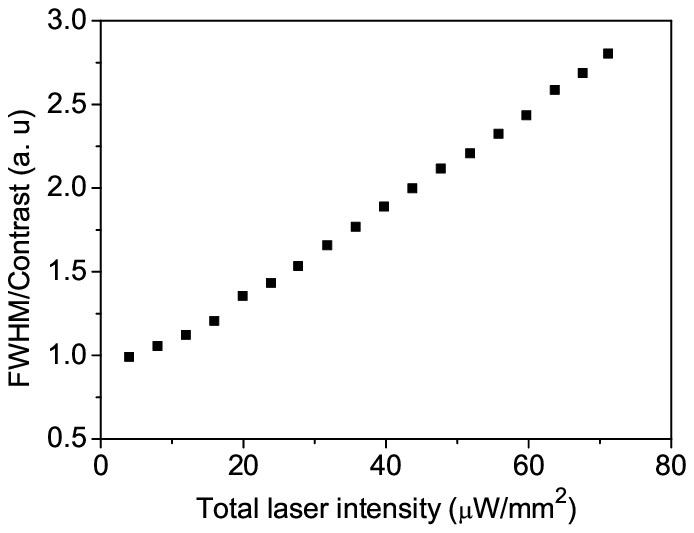}
\label{fig:stab-microcell}}
 \caption{(Color online) CPT resonance linewidth (a), CPT resonance contrast (b) and linewidth/contrast ratio (c) versus laser intensity for the cell 1. The laser beam diameter is 2 mm.}
\end{figure*}

\clearpage
\begin{figure*}
\centering
\subfigure[]{\includegraphics[width=0.4\linewidth]{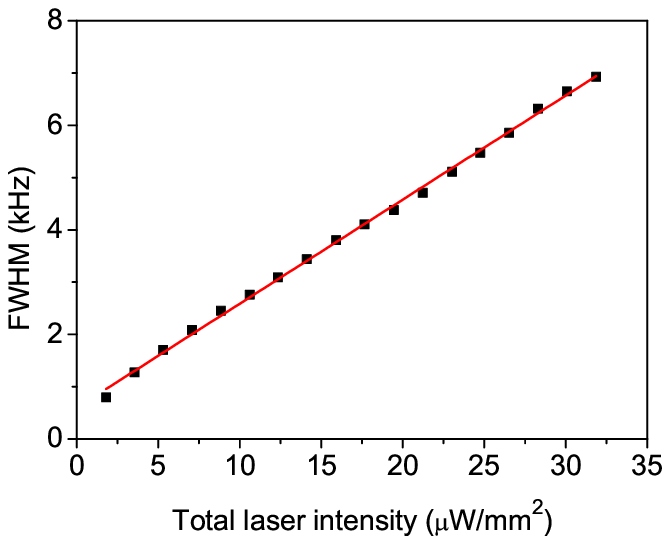}
\label{fig:largeur-midcell}} \vfill
\subfigure[]{\includegraphics[width=0.4\linewidth]{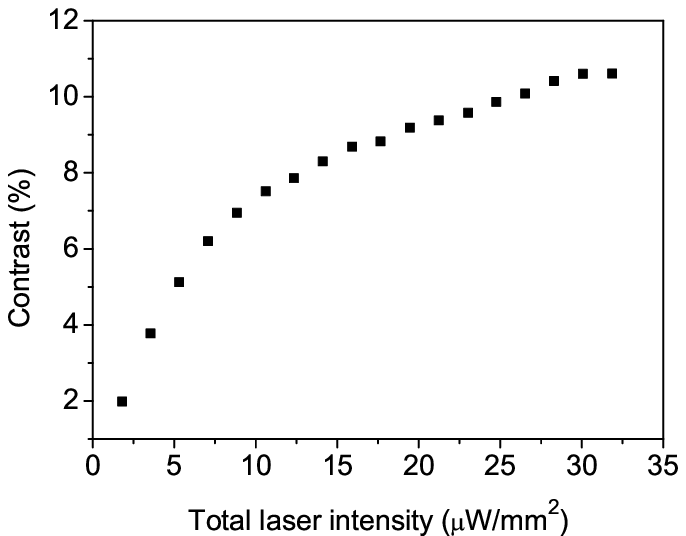}
\label{fig:C-midcell}}\vfill
\subfigure[]{\includegraphics[width=0.4\linewidth]{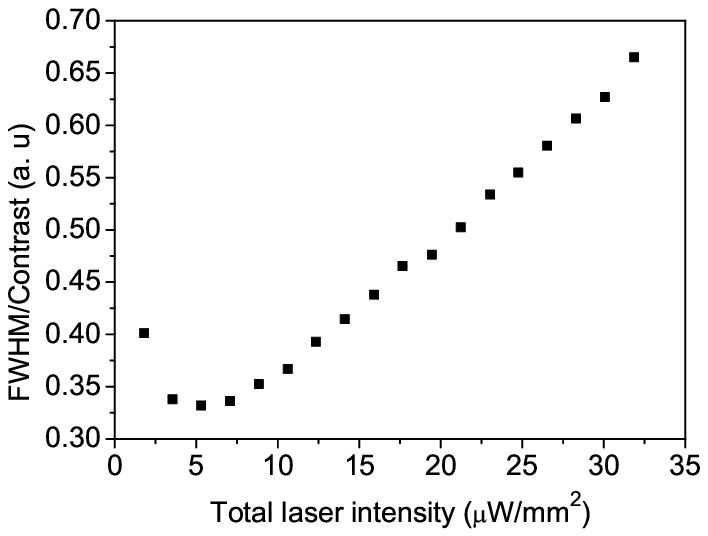}
\label{fig:stab-midcell}}
 \caption{(Color online) CPT resonance linewidth (a), CPT resonance contrast (b) and linewidth/contrast ratio (c) versus laser intensity for the cell 2. The laser beam diameter is 6 mm.}
\end{figure*}

\clearpage
\begin{figure*}
\centering
\subfigure[]{\includegraphics[width=0.4\linewidth]{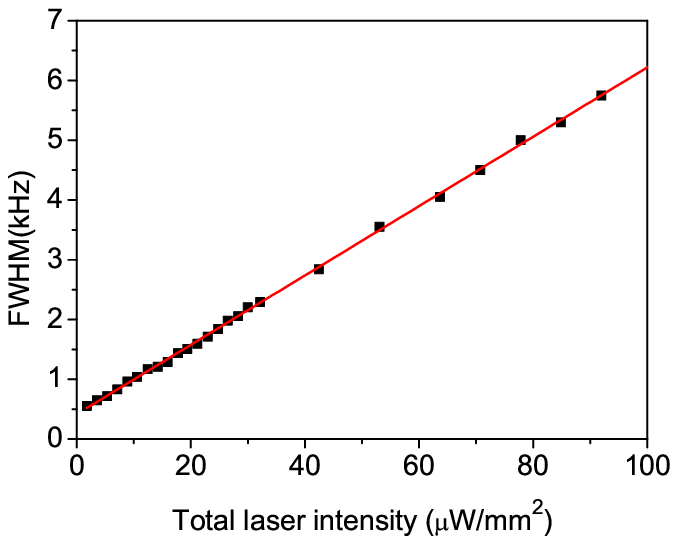}
\label{fig:largeur-bigcell}} \vfill
\subfigure[]{\includegraphics[width=0.4\linewidth]{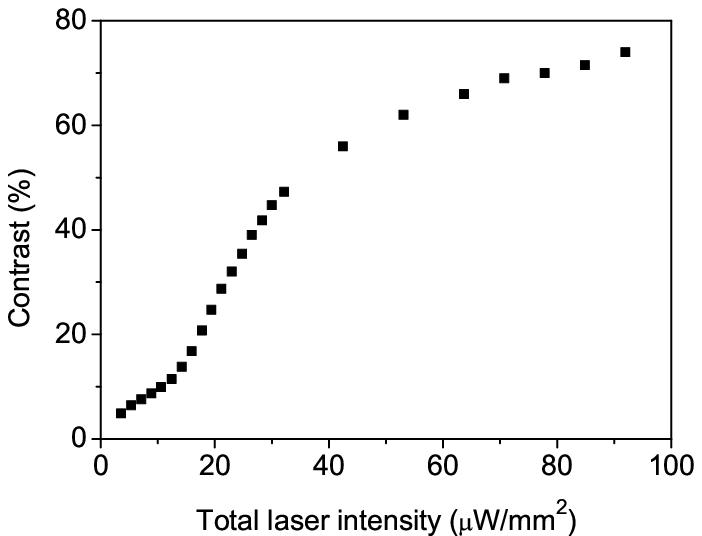}
\label{fig:C-bigcell}}\vfill
\subfigure[]{\includegraphics[width=0.4\linewidth]{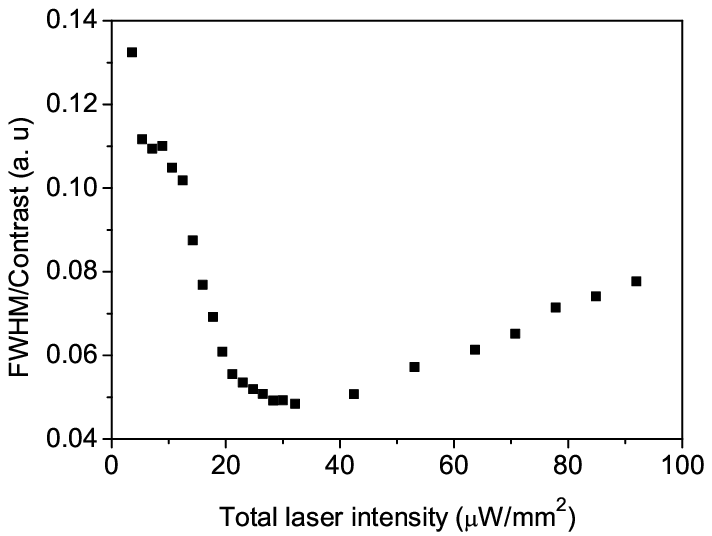}
\label{fig:stab-bigcell}}
 \caption{(Color online) CPT resonance linewidth (a), CPT resonance contrast (b) and linewidth/contrast ratio (c) versus laser intensity for the cell 3. The laser beam diameter is 6 mm.}
\end{figure*}

\clearpage
\begin{figure}[t!]
\centering
\includegraphics[width=\linewidth]{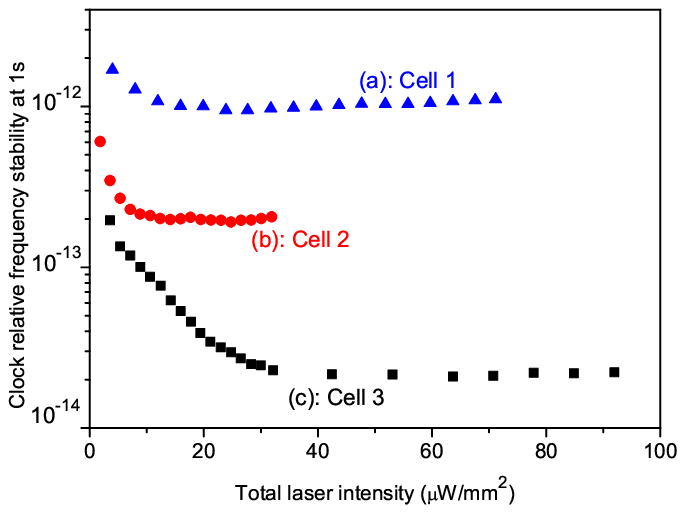}
\caption{(Color online) Estimated shot-noise limited clock short-term frequency stability at 1 s for the different cells with PPOP interaction. The combination of Eq. 1 and Eq. 2 is used for the calculation. The laser beam diameter is 2 mm for the cell 1 and 6 mm for cells 2 and 3. (a): Cell 1, (b): Cell 2, (c): Cell 3.} \label{fig:stab-bilan}
\end{figure}


\begin{thebibliography}{1}

\bibitem{Alzetta:1976}
G. Alzetta, A. Gozzini, L. Moi and G. Orriols, Nuovo Cimento B \bf{36}\rm, 5 (1976).

\bibitem{Arimondo:1996}
E. Arimondo, Prog. Opt., \bf{35}, \rm 257 (1996).

\bibitem{Fleischauer:RMP:05}
M. Fleischhauer, A. Imamoglu, and J. P. Marangos, Rev.
Mod. Phys. \bf{77}\rm, 633 (2005).

\bibitem{Wynands:1999}
R. Wynands and A. Nagel, Appl. Phys. B, \bf{20}\rm, 1-25 (1999).

\bibitem{Merimaa:2003}
M. Merimaa, T. Lindvall, I. Tittonnen and E. Ikonen, J. Opt. Soc. Am. B \bf{20} \rm, 273-279 (2003).

\bibitem{Nagel:EPL:1998}
A. Nagel, L. Graf, A. Naumov, E. Mariott, V. Biancalna, D. Meschede and R. Wynands, Europhys. Lett. \bf{44}\rm, 31-36 (1998).

\bibitem{Schwindt:APL:2004}
P. D. D. Schwindt, S. Knappe, V. Shah, L. Hollberg and J. Kitching, Appl. Phys. Lett. \bf{85}\rm, 26, 6409-6411 (2004).

\bibitem{Aspect:JOSAB:1989}
A. Aspect, E. Arimondo, R. Kaizer, N. Vansteenkiste and C. Cohen-Tannoudji, J. Opt. Soc. Am. B \bf{6}\rm, 2112-2124 (1989).

\bibitem{Bajcsy:Nature:2003}
M. Bajcsy, A. S. Zibrov and M. D. Lukin, Nature \bf{426}, \rm 638-641 (2003).

\bibitem{Vanier:APB:2005}
J. Vanier, Appl. Phys. B Lasers Opt., \bf{81}\rm 421-442 (2005).

\bibitem{Rutman}
J. Rutman and F. L. Walls, Proc. of the IEEE \bf{79}\rm, 953-960 (1991).

\bibitem{ShahKitching:2010}
V. Shah and J. Kitching
"Advances in Coherent Population Trapping for Atomic Clocks,"
Advances in Atomic, Molecular and Optical Physics, Elsevier, \bf{59}, \rm Chapter 2 (2010).

\bibitem{Jau:PRL:2004-1}
Y.-Y. Jau, A. B. Post, N. N. Kuzma, A.M. Braun, M.V. Romalis, and W. Happer, Phys. Rev. Lett. \bf{92}\rm, (11)~ 110801 (2004).

\bibitem{Jau:PRL:2004}
Y. Y. Jau, E. Miron, A. B. Post, N. N. Kuzma and W. Happer, Phys. Rev. Lett. \bf{93}\rm, (16)~ 160802 (2004).

\bibitem{Jau:PRL:2007}
Y. Y. Jau and W. Happer, Phys. Rev. Lett. \bf{99}\rm, 223001 (2007).

\bibitem{Zhang:OE:2012}
Y. Zhang, S. Qu and S. Gu, Opt. Exp. \bf{20}\rm, 6, 6400-6405 (2012).

\bibitem{Zanon:PRL:2005}
T. Zanon, S. Gu\'erandel, E. de Clercq, D. Holleville, N. Dimarcq
and A. Clairon, Phys. Rev. Lett. \bf{94}\rm, 193002 (2005).

\bibitem{Castagna:UFFC:2009}
N. Castagna, R. Boudot, S. Guerandel, E. De Clercq, N. Dimarcq and
A. Clairon, IEEE Trans. Ultrason. Ferroelec. Freq. Contr. \bf{56}\rm,
(2), 246-253 (2009).

\bibitem{Boudot:IM:2009}
R. Boudot, S. Guerandel, E. De Clercq, N. Dimarcq and
A. Clairon, IEEE Trans. Instr. Meas. \bf{58}\rm,
(4), 1217-1222 (2009).

\bibitem{Yim:RSI:2008}
S. H. Yim, T. H. Youn and D. Cho, Rev. Sci. Instr. \bf{79}\rm, 126104 (2008).

\bibitem{Yun:RSI:2011}
P. Yun, B. Tan, W. Deng and S. Gu, Rev. Sci. Instr. \bf{82}\rm, 123104 (2011).

\bibitem{Yun:RSI:2012}
P. Yun, B. Tan, W. Deng, J. Yang and S. Gu, Rev. Sci. Instr. \bf{83}\rm, 093111 (2012).

\bibitem{Taichenachev:JETP:2004}
A. V. Taichenachev, V. I. Yudin, V. L. Velichanslky, S. V. Kargapoltsev, R. Wynands, J. Kitching and L. Hollberg, {\em JETP Lett.} \bf{80}\rm, 477-481 (2004).

\bibitem{Taichenachev:JETP:2005}
A. V. Taichenachev, V. I. Yudin, V. L. Velichanslky and S. A. Zibrov, {\em JETP Lett.} \bf{82}\rm, 398-403 (2005).

\bibitem{Mikhailov:JOSAB:2010}
E. E. Mikhailov, T. Horrom, N. Belcher, and I. Novikova, J. Opt. Soc. Am. B \bf{27}\rm, 417-422 (2010).

\bibitem{Zibrov:PRA:2010}
S. A. Zibrov, I. Novikova, D. F. Phillips, R. L. Walsworth, A. S. Zibrov,
V. L. Velichansky, A. V. Taichenachev, and V. I. Yudin, Phys. Rev. A \bf{81}\rm, 013833 (2010).

\bibitem{Watabe:AO:2009}
K. Watabe, T. Ikegami, A. Takamizawa, S. Yanagimachi, S. Ohshima and S. Knappe, Appl. Opt. \bf{48}\rm, (6)~ 1098-1103 (2009).

\bibitem{a}
$E_r$ is right-handed circularly polarized from the detector point of view, left-handed circularly polarized from the source point of view. We note $\sigma^-$ when $\overrightarrow{B}$ is oriented along O$z$, $\sigma^+$ for the opposite direction. All senses are inverted for $E_l$.

\bibitem{VA}
J. Vanier and C. Audoin, "The quantum physics of atomic frequency standards", Adam Hilger, Bristol (1989).

\bibitem{Cohen:1991} C. Cohen-Tanoudji, \emph{Cours Collège de France}, 1991-1992,
\href{ici}{http://www.phys.ens.fr/cours/college-de-france/1991-92/cours4/cours4.pdf}.

\bibitem{Stahler:OL:2002}
M. St\"{a}hler, R. Wynands, S. Knappe, J. Kitching, L. Hollberg, A. Taichenachev, and V. Yudin, Opt. Lett., \bf{27}\rm, 1472 (2002).

\bibitem{Knappe:2000}
 S. Knappe, W Kemp, C. Affolderbach, A; Nagel, and R. Wynands, Phys. Rev. A,  \bf{61}\rm, 12508 (2000).

\bibitem{Korsunsky:1999} E. A. Korsunsky and D. V. Kosachiov, Phys. Rev. A,  \bf{60}\rm, 4996, (1999).

\bibitem{Gornyi:1989} M. B. Gornyi, B. G. Matisov and Yu. V. Rozhdestvenskii, Sov. Phys. JETP, \bf{68}\rm, 728, (1989).

\bibitem{Liu:IM:2012}
X. Liu and R. Boudot, IEEE Trans. Instr. Meas. \bf{61}\rm, 10, 2852-2855 (2012).

\bibitem{Boudot:IM:NLTL:2009}
R. Boudot, S. Gu\'erandel, E. De Clercq, N. Dimarcq and A. Clairon, IEEE Trans. Instr. Meas. \bf{58}\rm, 3659-3665 (2009).

\bibitem{Rubiola:xtal}
E. Rubiola and V. Giordano, IEEE Ultrason. Ferroelec. Freq. Contr., \bf{54}\rm, 1, 15-22 (2007).

\bibitem{Becker:APL:85}
R. A. Becker and R. C. Williamson, Appl. Phys. Lett. \bf{47}\rm, 1024-1026 (1985).

\bibitem{Harvey:JLT:98}
G. T. Harvey, Journ. Light. Tech. \bf{6}\rm, 872-876 (1998).

\bibitem{Boudot:RSI:2005}
R. Boudot, C. Rocher, N. Bazin, S. Galliou and V. Giordano, Rev. Sci. Instr., \bf{76}\rm, 095110 (2005).

\bibitem{Liu:OL:2012}
F. Liu, C. Wang, L. Li, L. Chen, Opt. Laser Tech., \bf{45}\rm, 775 (2012).
\bibitem{ZhuCutler:2000}
M. Zhu and L. S. Cutler, Proc. 32$^{th}$ Annual Precise Time and Time Interval (PTTI) Meeting, Reston, VA, USA, pp. 311-323 (2000).

\bibitem{Shah:APL:2006}
V. Shah, V. Gerginov, P. D. D. Schwindt, S. Knappe, L. Hollberg and
J. Kitching, Appl. Phys. Lett. \bf{89}\rm, 151124 (2006).

\bibitem{Happer:APL:2009}
B. H. McGuyer, Y. Y. Jau and W. Happer, Appl. Phys. Lett. \bf{94}\rm, 251110 (2009).

\bibitem{Miletic:APB:2012}
D. Miletic, C. Affolderbach, M. Hasegawa, R. Boudot, C. Gorecki and G. Mileti, Appl. Phys. B \bf{109}\rm, 89 (2012).

\bibitem{Hasegawa:SA:2011}
M. Hasegawa, R. K. Chutani, C. Gorecki, R. Boudot, P. Dziuban, V. Giordano, S. Clatot, L. Mauri, Sensors Actuat. A: Phys. \bf{167}\rm, 594 (2011).

\bibitem{Boudot:JAP:2011}
R. Boudot, P. Dziuban, M. Hasegawa, R. K. Chutani, S. Galliou, V. Giordano and C. Gorecki, Journ. Appl. Phys. \bf{109}\rm, 014912 (2011).


\bibitem{Knappe:JOSAB:2001}
S. Knappe, R. Wynands, J. Kitching, H. G. Robinson and L. Hollberg, Journ. Opt. Soc. Am. B \bf{18}\rm, 1545-1553 (2001).

\bibitem{Micalizio:Metrologia:2012}
S. Micalizio, C. Calosso, A. Godone and F. Levi, Metrologia \bf{49}\rm, 425-436 (2012).

\bibitem{Knappe:Elsevier:2010}
S. Knappe, MEMS Atomic Clocks, Comprehensive Microsystems, Elsevier B.V \bf{3} \rm, 571-612 (2010).

\bibitem{Dicke:PR:1953}
R. H. Dicke, Phys. Rev. \bf{89}\rm, 472-473 (1953).

\bibitem{Zanon:PRA:2011}
T. Zanon, E. De Clercq and E. Arimondo, Phys. Rev. A \bf{84}\rm, 062502 (2011).

\bibitem{Yun:EPL:2011}
P. Yun, Y. Zhang, G. Liu, W. Dang, L. Yu, S. Gu, EPL \bf{97}\rm, 63004 (2012).





\end{thebibliography}
\end{document}